\documentclass{aa}
\usepackage{graphicx}
\usepackage{txfonts}
\usepackage{natbib}

\begin{document}
   \title{An updated survey of globular clusters in M 31}
\subtitle{I. Classification and radial velocity for 76 candidate clusters 
\thanks{Table 5 is only available in electronic form at the CDS via anonymous
ftp to cdsarc.u-strasbg.fr (130.79.128.5) or 
via http://cdsweb.u-strasbg.fr/cgi-bin/qcat?J/A+A/. 
}
 }

   \author{S. Galleti\inst{1,2}, L. Federici\inst{2}, M. Bellazzini\inst{2},
           A. Buzzoni\inst{2},
          \and
       F. Fusi Pecci\inst{2}
\thanks{Based on observations made at La Palma, at the Spanish Observatorio del
 Roque de los Muchachos of the IAC,
with the Italian Telescopio Nazionale Galileo (TNG) operated by the Fundaci\'on
Galileo Galilei of INAF and with the William Herschel Telescope of the Isaac Newton Group,
on TNG-ING sharing-time agreement. Also based on observations made with the G.B.\ Cassini
Telescope at Loiano (Italy), operated by the Osservatorio Astronomico di Bologna (INAF).}
          }

   \offprints{S. Galleti}

   \institute{Universit\`a di Bologna, Dipartimento di Astronomia
             Via Ranzani 1, 40127 Bologna, Italy\\
            \email{silvia.galleti2@unibo.it}
         \and
             INAF - Osservatorio Astronomico di Bologna,
              Via Ranzani 1, 40127 Bologna, Italy\\
          \email{michele.bellazzini@oabo.inaf.it, luciana.federici@oabo.inaf.it, \\
          alberto.buzzoni@oabo.inaf.it, flavio.fusipecci@oabo.inaf.it} }

     \authorrunning{S. Galleti et al.}
   \titlerunning{A large survey of globular clusters in M31. I.
   Classification and radial velocity.}

   \date{Received 29 March 2006 / Accepted 18 May 2006}

\abstract
{}
{We present the first results of a large spectroscopic survey of
globular clusters and candidate globular clusters in the nearby M~31
galaxy. The survey is aimed at the classification of known candidate
M~31 clusters and at the study of their kinematic properties.}
{We obtained low-resolution spectroscopy ($\lambda/\Delta\lambda \simeq 800
   - 1300$) for 133 targets, including 76 yet-to-confirm candidate clusters
   (i.e.\ with no previous spectroscopic information), 55 already-confirmed
   genuine M~31 clusters, and 2 uncertain candidates. Our observations
   allowed a reliable estimate of the target radial velocity, within
   a typical accuracy of $\sim \pm 20$~km~s$^{-1}$.
   The observed candidates have been robustly
   classified according to their radial velocity and shape parameters
   that allowed us to confidently discriminate between point sources and extended
   objects even from low-spatial-resolution imagery.}
{In our set of 76 candidates clusters we found:
   42 newly-confirmed bona-fide M~31 clusters, 12 background galaxies, 17
   foreground Galactic stars, 2 H{\sc ii} regions belonging to M~31 and 3 unclassified
   (possibly M~31 clusters or foreground stars) objects.
   The classification of a few other candidates not included in our survey has
   been also reassessed on various observational bases.
All the sources of radial velocity estimates
   for M~31 known globular clusters available in the literature have been compared and
   checked, and a homogeneous general list has been obtained for 349 confirmed
   clusters with radial velocity.}
{Our results suggest that a significant number of genuine clusters ($\gtrsim 100$)
is still hidden among the plethora of known candidates proposed by
   various authors. Hence our knowledge of the globular cluster system of
   the M31 galaxy is still far from complete even in terms of simple
   membership.}


   \maketitle
%

\section{Introduction}

Since the dawn of extragalactic astronomy (Hubble \cite{hubble}),
the study of the globular cluster (GC) system around the Andromeda
galaxy (M31) stand out as an important field of investigation,
providing the ideal counterpart to compare with the GC system of our
own Galaxy and a testbed for observational techniques to be applied
to GC systems of more distant galaxies. In present days it is
possible to compare directly integrated properties and resolved
Color Magnitude Diagrams (CMD) of M31 clusters, as recently attained
with the Hubble Space Telescope (HST) (see, e.g. Fusi Pecci et al.
\cite{fp96}, Brown et al. \cite{brown04}, Rich et al. \cite{richH}).

In spite of a similar mass and morphology, M31 is recognized to display several differences
in its stellar and clusters content with respect to the MW. For instance, the halo stellar
populations of the two galaxies (the typical environment of most GCs) widely differ
in average metallicity; while stars in the MW halo are predominantly metal poor
($\langle [Fe/H]\rangle \simeq -1.6$~dex  (Laird et al. \cite{laird}), those in the outer
regions of M31 have $\langle [Fe/H]\rangle \simeq -0.5$~dex
(see, for example, Holland et al. \cite{holland}; Bellazzini et al. \cite{michele};
Rich et al. \cite{rich}; Durrell et al. \cite{durrell}, and references therein).
The GC system of M31 is much more populous than that of the MW, with more
than 300 confirmed clusters, to compare with the $\sim$150 of the MW.
In particular, Barmby et al. \cite{barm01} estimated the total number of M31
GCs to be $475 \pm 25$, i.e.\ more than a factor of three larger than in the Milky Way.
It is quite clear that understanding the reasons of such striking differences
may shed light on the formation histories of the two galaxies and, at large,
on the general process of galaxy formation.

Since early systematic surveys (see, among others, Vete\v{s}nik
\cite{vet62}, van den Bergh \cite{syd67,syd69}, Baade \& Arp
\cite{BA}, Sargent et al. \cite{sg77}, Crampton et al. \cite{ccsc85}
Battistini et al. \cite{bat80,bat82,bat87,bat93}, Sharov et al.
\cite{sha95}, Mochejska et al. \cite{Mochejska}, Barmby et al.
\cite{barm00}, and references therein), most of the known M~31 GC
candidates (hereafter GCCs) have been typically identified by visual
inspection of wide-field photographic plates, and a more exhaustive
analysis with CCD cameras is still partially lacking. On this line,
a few recent studies (see, for example, Racine \cite{racine91},
Barmby \& Huchra \cite{barm01}, Mochejska et al. \cite{Mochejska},
Huxor et al. \cite{hux,huxext}, Barmby et al. \cite{barm01}, Perrett
et al. \cite{perr_cat}, Fusi Pecci et al. \cite{ffp}, Beasley et al.
\cite{beas}, Puzia et al. \cite{puzia}) have clearly shown that this
kind of new-generation surveys may significantly change our
knowledge of the number and nature of the clusters harbored in the
M~31 system. To date, there are still hundreds of known GCCs
($\sim$700, see Galleti et al. \cite{silvia}, henceforth G04) whose
nature remains to be ascertained. 

\begin{figure} \centering \includegraphics[width=\hsize]{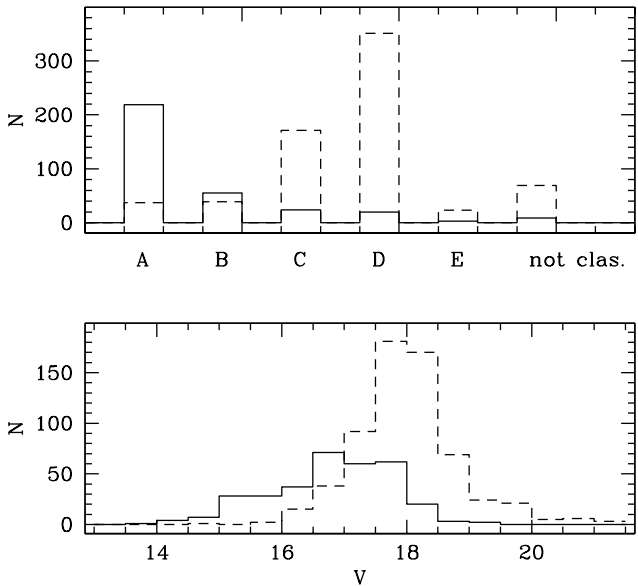}
   \caption{Distribution of the original quality-class classifications of
   Battistini et al.  (upper panel), and the V luminosity function  (lower
   panel) for confirmed M~31  clusters (that is all $c = 1$ entries in the RBC
   -- see Sect.~1 -- according to G04; solid histogram) and candidate M~31
   clusters ($c = 2$ RBC flag; dashed histogram).  Note the huge number of
   yet-to-confirm candidates at $V>17.0$, mainly belonging to class C and D of
   the Battistini et al. classification scale. }
   \label{faint}
   \end{figure}

Most of these candidates pertain
to the faint-end tail of the GC luminosity function which, at
present, is far from complete (see Fig.~\ref{faint}).
Battistini et al. \cite{bat80,bat82,bat87,bat93}, ranked the
majority of cluster candidates according to a quality class (from
``A'' to ``E'' in the sense of decreasing confidence level) related
to their appearance on the original images. The upper panel of
Fig.~\ref{faint} shows that the majority of class A and B candidates
have already been confirmed (either spectroscopically or, in a few
cases, by high resolution images), while most of class
C and D targets remain to be explored. In any case, one has to
consider that any comprehensive sample of M~31 GCs should forcedly
rely on coarse databases whose homogeneity is difficult to assess.
On the other hand, a catalog of M31 clusters as complete and homogeneous 
as possible is clearly overdue, and this is a fundamental step for any 
meaningful comparison with other systems. In this framework we have 
started a long-term project to step ahead with this ambitious goal.


In G04 we re-analyzed the photometric and classification data
available in the literature, reporting coarse photometry to a
self-consistent CCD-based magnitude scale, and providing infrared
information for several hundred GCCs as well, with J,H,K magnitudes
from the 2MASS database (Skrutskie et al. \cite{Skrutskie}). The resulting
catalog (Revised Bologna Catalog, hereafter RBC\footnote{The catalog is available
electronically at {\tt http://www.bo.astro.it/M31/}.}) contains at present 1164 entries,
including all the already confirmed clusters, the known but
yet-to-confirm candidates, and the proposed candidates whose
``non-M31-GC'' nature has been definitely established. We believe
that keeping record of the latter objects is very useful all the
way, not only to avoid duplicating observations, but also to
characterize the observing properties of the typical contaminants,
in order to tune up selection criteria and maximize the return of
future observational campaigns (see G04 for details and discussion).

For reader's better convenience, it may be useful to recall again the RBC
classification scheme, as originally used in G04. The classification flag $c$
can therefore assume the following values: $c=1$
confirmed clusters,  $c=2$ candidate clusters, $c=3$ uncertain candidates, $c=4$
background galaxies,  $c=5$ H{\sc ii} regions, $c=6$ foreground stars. In addition,
as a result of the present discussion (see Sect.~4.3, below), we will  also introduce here a
new class ($c = 7$) including asterisms/associations.

As a further line of attack in our work, we are carrying on a survey of GCCs
located at large (projected) distance from the center of M~31, a nearly
unexplored realm. A first pilot run of the survey led to the discovery of B514 (Galleti et al.
\cite{b514}, hereafter G05), the outermost cluster  of the Andromeda galaxy ever known.

Finally we have undertaken a large spectroscopic follow up of known
candidates, to assess their nature and  study their physical
properties. This is the subject of the present paper, where we deal
with  classification and kinematic of 76 M~31 GCCs never surveyed
before.

In the following Sec.~2 we briefly introduce the problem of M~31 GCCs classification,
and describe the observational material and its reduction procedures. Sect.~3 is
devoted to radial-velocity estimates for our sample and the comparison of our
results with other spectroscopic datasets available in the literature. Sect.~4 deals
with the classification of the observed candidates and in Sect. 5 we use the kinematical
information for the updated sample of confirmed clusters to estimate the mass of M~31.
Finally, in Sec.~6 we summarize and discuss our results.

In a companion paper (Galleti et al., in preparation) we will complete our study
by assessing the problem of a self-consistent metallicity scale for M~31 GCs,
relying on homogeneous measures of the Lick indices (Trager et al. \cite{trager}).

\section{Observations and data reduction}

A spectroscopic database of 133 targets has been collected through
different observing runs along 2004/05, carried out at the Roque de los Muchachos
(La Palma, Spain) observing
facilities of the 3.5m Telescopio Nazionale Galileo (TNG) and the 4.2m William Herschel
telescope (WHT), and at the  ``G.B. Cassini'' 1.52m telescope of the Loiano Observatory (Italy).
Our sample comprises 76 M~31 GCCs, while further 55 confirmed clusters ($c = 1$ RBC flag)
were included to consistently match our results with other external data sources in the
literature, together with 2 more questioned objects that need a more definitive spectroscopic
assessment. The survey mainly aims at achieving accurate radial-velocity measurements
for a wide sample of M~31 GCCs through low-resolution long-slit and fiber spectroscopy.

The kinematical piece of information alone can easily discriminate between genuine M~31 GCs
and the most relevant class of spurious contaminants, i.e.\ background galaxies.
The nature of the latter sources can in fact be recognized because of their cosmological
recession velocities (typically $6000 \lesssim cz \lesssim 50\,000$~km~s$^{-1}$, see G04),
much larger (and opposite in sign) than the systemic velocity of M31
(namely $V_r \simeq -301$~km/s, Van den Bergh \cite{sydbook}).

On the other hand, radial velocity alone may not be sufficient to discriminate between  M~31 GCs
and foreground interlopers -- mostly MW stars -- the other major source of contamination
for our sample. In these cases  one has to recur to morphological criteria in order to
confidently discriminate between point (i.e.\ stars) and extended sources (GCs). To assess this
important point, we also complemented our spectroscopic analysis with supplementary imagery,
taken at the Loiano telescope, for an independent but largely overlapping sample of 86 RBC objects.
We will return on Sec.\ 4 for a full discussion of these results.

\subsection{WYFFOS Data}

Most of our sample has been observed during the nights of Nov~21 and 22 2004, using the
AutoFib2+Wide Field Fibre Optic Spectrograph (WYFFOS; Telting \& Corradi
\cite{manaf2}), mounted at the WHT prime focus.
We adopted the R1200B grating, covering the $\lambda\lambda = 4000 \to 5700$~\AA\
spectral range with a FWHM wavelength resolution of 2.2~\AA.
WYFFOS is equipped with a mosaic of two EEV-42-80 CCDs, windowed and mosaiced such as to
have $\simeq 4300 \times 4200$~px$^2$ in total,
that were read in $2 \times 2$ binning mode, with a scale of 0.4~\AA/pixel.
We used the Small Fibre module, which is made of 150 1.6~arcsec science fibres,
and 10 fiducial bundles for target acquisition and guiding. Four different fibre configurations
were set up in order to target a total of 183 confirmed clusters and GCCs.

A total of 10 science exposures ranging from 1200 to 2700~sec were acquired
(four in the first night and six in the second night).
Calibrating exposures included bias, sky and lamp flat-field, and He/Ne
comparison lamps.
For each of the four pointings we typically allocated a fraction of $\sim$30\%
of the whole fibre configuration to
simultaneously sample the sky level. The sky emission was also probed before and after each
target exposure by dithering the same fibre configuration.
Data were processed in IRAF, using a dedicated
package ({\sf WyffosREDUC}) written by Pierre Leisy. The code was modified by S.G. to
automatically account for the whole data-reduction pipe-line; the procedure has been
extensively tested, comparing the results with those obtained with standard
reduction techniques for long-slit spectra.
The {\sf WyffosREDUC} package is based on the IRAF task {\sf DOFIBER}, and corrects frames for bias,
flat-field and for fiber throughputs, extracts spectra for each fibre, calibrates
them in wavelength, and performs sky subtraction as described below.
Fibre throughputs have been obtained from exposures of {\em sky flat-field} frames.


   \begin{figure} \centering \includegraphics[width=0.8\hsize]{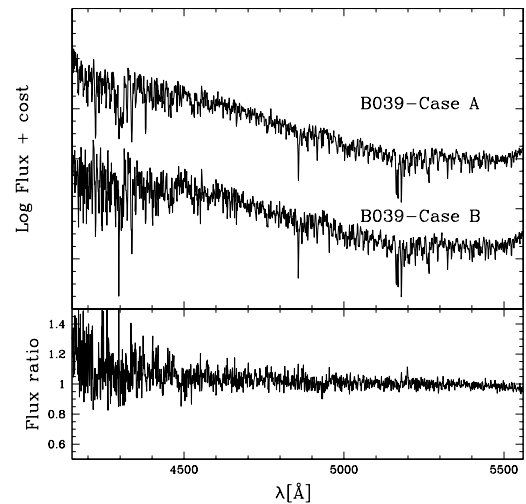}
   \caption{  An illustrative example of the impact of different sky-subtraction procedures
   on WYFFOS spectroscopic observations of cluster B039 {\it (upper pane)}. The raw instrumental spectrum
   is displayed after sky subtraction {\it i)} according to  a ``master spectrum'' (Case A)
   homogeneously sampled across the whole field of view by $\sim35$ suitably allocated fibres,
   and {\it ii)} by evaluating sky level from $20\arcsec$ off-target dithered images
   taken with the scientific fiber configuration just before and after the on-target shot (Case B).
   An arbitrary offset in $\log Flux$ has been added to the data for graphical optimization.\protect\\
   {\it Lower panel} - The flux ratio of Case A and B raw spectra. Global standard deviation (per pixel element) amount to a 9\% (with no
   evident drift with wavelength), improving to a 4\% scatter if we restrain 
   to $\lambda \ge 4500$~\AA.
}
   \label{sky}
   \end{figure}

 Sky correction needs special attention when dealing with
fiber observations; in this regard we explored two independent techniques:
{\it i)} a master sky spectrum was mapped by a grid of $\sim$35 fibers homogeneously
distributed across the whole field of view. This output was then subtracted to
each science spectrum after rescaling for the appropriate fiber throughput;
{\it ii)} the on-target shots were sided (before and after) by two off-target images
(dithering the telescope $20\arcsec$ E away) to sample sky level with the same fiber configuration.
No substantial differences were found between the two procedures (see Fig.~\ref{sky}
for an illustrative example), and we eventually decided to apply  option {\it (i)} for our analysis.

Two M~31 globular clusters (namely B225 and B158), for which very accurate radial
velocity estimates are available in the literature (Dubath \&
Grillmair \cite{DG}), were observed in two different pointings and used as
reference templates to set radial-velocity zero point. These reference spectra reach
$S/N> 50$ per pixel, to be compared with a typical $S/N\simeq 10$ for the GCC spectra. Two
exposures for each scientific target have been acquired. The same strategy has
been adopted also for the runs described below.

The poor seeing conditions (FWHM~$\sim 1.5\arcsec$), in addition to thin-cloud
sky coverage during both observing nights, and the less-than-perfect positioning of some
fibres eventually limited the final signal-to-noise  of several faint targets.
For this reason only 116 out of the 183 targets ended up with a useful spectrum for our
analysis.

\subsection{BFOSC Data}

Long-slit spectra for 8 relatively bright ($V\le 16.5$) targets were obtained with the
low-resolution spectrograph BFOSC (Gualandi \& Merighi \cite{manbfosc})
operated at the Loiano Observatory. The detector was a thinned, back illuminated EEV CCD,
with $1300 \times 1340$~px$^2$. The observations were carried out
under typical seeing conditions (FWHM~$\sim 1.5 \arcsec$), on
six nights during 2004 (Sep 18-19, Nov 16-17) and 2005 (Jan 02-05). With a
1.5 arcsec slit, the adopted setup  provided a FWHM spectral
resolution $\Delta \lambda = 4.1$~\AA\  (${\lambda}/{\Delta \lambda} \simeq
1300$) covering the range $\lambda\lambda = 4200 \to 6600$~\AA. We took a He-Ar
calibration-lamp spectrum after each scientific exposure, maintaining the same
telescope pointing. Exposure times were typically 45 minutes, yielding spectra
with a characteristic $S/N \simeq 8$ per resolution element.

During each observing night and with the same instrumental setup, we also collected
accurate ($S/N > 70$) observations of four radial-velocity template targets, namely the
same M~31 GCs adopted for the WYFFOS run (B158 and B225) plus stars HD\,12029 and HD\,23169
(heliocentric radial velocities from the SIMBAD database).
Bias, flat field, and sky subtraction were carried out using
standard packages in IRAF, as described in Galleti et al. \cite{b514}.

\subsection{DOLORES Data}

The imager/spectrograph DoLoRes at the TNG was used in the nights of Sep~8 and Oct~8, 2004,
to acquire long-slit spectra of 9 M~31 GCCs.
DoLoRes is equipped with a $2048 \times 2048$~px$^2$ thinned and back-illuminated Loral CCD
array providing a $9.4\arcmin \times 9.4\arcmin$ field of view.
The adopted MRB grism yielded a resolution of 6~\AA\ FWHM ($\lambda/\Delta \lambda = 875$)
with a $1\arcsec$ slit, across a $\lambda\lambda = 3800 \to 6800$~\AA\ spectral range.
We typically exposed 10-15 minutes, reaching $S/N \simeq 13$
per resolution element. A He-lamp spectrum was acquired after each science frame for
wavelength calibration. Again, during each night we obtained good ($S/N \gg 50$) spectra
for the template clusters B158 and B225. The data reduction procedure was the same as for
the BFOSC data.

\section{Radial velocities}

The heliocentric radial velocities ($V_r$) of GCCs
were obtained by cross-correlation  with the templates spectra,
using the IRAF/{\sf fxcor} package (see Tonry \& Davis  \cite{tonry} for
details of the technique).
We applied a square filter to dampen the highest and lowest frequency Fourier
components, that heavily masked the narrow peaks in the power spectrum.
We then fitted the power peaks with Gaussians.
The typical internal velocity errors on a single measure were $\sim$30~km~s$^{-1}$
for WYFFOS, $\sim$50~km~s$^{-1}$ for BFOSC and  $\sim$ 65~km~s$^{-1}$ for DoLoRes
spectra.\footnote{The standard cross correlation procedure provided poor results
for background galaxies, given the reduced wavelength range in common with the
reference template globulars and, in most cases, the presence of strong emission
lines.  For those targets with $V_r>+6000$~km~s$^{-1}$ we therefore derived the value
of $cz$ directly from the measure of a few strong emission/absorption spectral features
like the [OII]3727, [OIII]5007, Ca HK, H$\beta$, H$\delta$, H$\gamma$, and MgH lines.
This led to sensibly higher (but still fully acceptable) uncertainties on the
inferred value of $cz$ in Table~\ref{vr_sur}.}

For each target we cross-correlated two independent spectra with (at least) four
template spectra, obtaining $\ge 8$ semi-independent estimates of radial velocity.
We take the average of these values as our final $V_r$, and the standard
deviation as our final uncertainty on $V_r$. The typical uncertainties
are $\sim$14~km~s$^{-1}$ for WYFFOS data, $\sim$19~km~s$^{-1}$ for BFOSC, and
$\sim$33~km~s$^{-1}$ for DoLoRes data.

At the end of the analysis we obtained reliable $V_r$ estimates for 133
targets (116 targets from WYFFOS data, 9 from DoLores data, and 8 from BFOSC data),
76 of which were previously unconfirmed GCCs. Of the latter, 12 objects were
eventually classified as background galaxies due to their evident cosmological redshift
and, in 9 cases, also to striking line emission in their spectra (see Sect.~4.2 for
further details). In addition, two
more objects, namely G137 and G270, display clear $H\beta$, $H\gamma$ and [O{\sc iii}]
emission lines and a value of $V_r$ compatible with M~31 H{\sc ii} regions
(see, for a comparison, Diaz et al. \cite{diaz}).

The list of the observed targets is reported in Table~\ref{vr_sur}, together
with their apparent V magnitude and V-K color (cols.\ 2 and 3, from G04),
the estimated $V_r$ and the associated uncertainty (cols.\ 4 and 5), the original
quality class (col.\ 6), according to Battistini et al. \cite{bat82,bat87}), and
the instrument used to obtain the spectra (col.\ 7).

\begin{table*}[!ht]
\scriptsize
\caption{Radial velocities of observed GCCs and GCs.}             
\label{vr_sur}      
\begin{tabular}{l c c r@{$~~\pm$} r c l | l c c r@{$~~\pm$} r c l}
\hline
        &    &     &   \multicolumn{2}{c}{}   &     &         &      &    &     &   \multicolumn{2}{c}{}    &     &      \\
Name    & V  & V-K & $V_r$ & ~~$\sigma$($V_r$)& qc$^{(\dagger)}$ & Instrum.      &   Name    & V  & V-K & $V_r$ & ~~$\sigma$($V_r$)& qc$^{(\dagger)}$ & Instrum.     \\   
        &    &     &  \multicolumn{2}{c}{~~~[km~s$^{-1}$]}   &     &      &      &    &     &  \multicolumn{2}{c}{~~~[km~s$^{-1}$]}   &     &        \\
\hline
        &    &     &    \multicolumn{2}{c}{}   &     &         &     &    &     &    \multicolumn{2}{c}{}     &     &        \\
\multicolumn{7}{c}{\hrulefill ~~~Previously unconfirmed Globular Cluster Candidates~~~ \hrulefill} &    NB70  & 14.89 &   2.47  &   --17   &  11  & E  & BFOSC \\
  B003  & 17.57 &   2.03  &   --351  &    11  & A  & WYFFOS  &    SH01(*)  & 15.82 &      &   21500  &  600   &     & BFOSC   \\
  B022  & 17.36 &   1.85  &   --407  &    14  & A  & WYFFOS  &   B295D  & 17.86 &   3.69  &   32600  &  600   & D  & WYFFOS   \\
  B032  & 17.61 &   3.55  &   --516  &     8  & B  & WYFFOS  &   B330D  & 15.99 &   1.59  &    --62  &  18    & D  & BFOSC      \\
  B060  & 16.75 &   2.61  &   --484  &    25  & A  & WYFFOS  &    BA21(*)  & 16.64 & 2.77  &  36600  &  1400  &   & BFOSC      \\
  B067  & 17.25 &   1.99  &   --377  &    10  & A  & DoLoRes     &    G137(*)  & 17.81 &      &   --256  &  28    &     & WYFFOS      \\
  B070  & 17.07 &   2.25  &   --301  &    15  & A  & WYFFOS  &    G270(*)  & 17.30 &      &    --72  &  23      &    & WYFFOS     \\
  B071  & 17.79 &         &   --479  &     8  & B  & WYFFOS  &    H126  & 16.76 &   1.77  &    --19  &  20    &  & WYFFOS       \\
  B077  & 17.50 &   2.94  &   --681  &    25  & A  & DoLoRes     &    M049  & 18.26 &   2.66  &   --247  &  17    & D  & WYFFOS     \\
  B087  & 17.93 &         &   --382  &    16  & B  & WYFFOS  &\multicolumn{7}{c}{\hrulefill ~~~Previously observed controversial objects~~~ \hrulefill} \\
  B099  & 16.74 &   3.15  &   --200  &    11  & A  & WYFFOS  &    B341  & 16.37 &   2.84  &   --352  &  32  & A  & BFOSC     \\
  B100  & 17.91 &   3.22  &   --376  &    25  & B  & WYFFOS  &    B409  & 12.53 &   2.00  &   --39   &  10  & A  & BFOSC     \\
  B111  & 16.80 &   2.27  &   --414  &    10  & A  & WYFFOS  &\multicolumn{7}{c}{\hrulefill ~~~Previously confirmed Globular Clusters~~~ \hrulefill} \\
  B155  & 17.97 &   3.25  &   --401  &    34  & B  & WYFFOS  &    B006  & 15.50 &   2.87  &   --228  &  14  & A  & WYFFOS     \\
  B162  & 17.48 &   3.65  &   --146  &     8  & A  & WYFFOS  &    B012  & 15.09 &   2.35  &   --364  &  3 & A  & WYFFOS      \\
  B168  & 17.63 &   4.11  &   --190  &    14  & B  & WYFFOS  &    B017  & 15.95 &   3.36  &   --505  &  10  & A  & WYFFOS     \\
  B169  & 17.08 &   3.14  &   --177  &     7  & A  & WYFFOS  &    B019  & 14.93 &   2.95  &   --221  &  7  & A  & WYFFOS      \\
  B187  & 17.17 &   2.85  &   --130  &    10  & A  & WYFFOS  &    B034  & 15.47 &   2.97  &   --537  &  8  & A  & WYFFOS      \\
  B189  & 16.99 &   3.34  &   --148  &     8  & A  & WYFFOS  &    B039  & 15.98 &   3.60  &   --242  &  16  & A  & WYFFOS     \\
  B194  & 17.19 &   2.08  &   --354  &    21  & A  & WYFFOS  &    B042  & 16.29 &   4.13  &   --279  &  26  & A  & WYFFOS     \\
  B215  & 17.13 &   2.90  &   --164  &     3  & A  & WYFFOS  &    B051  & 16.08 &   3.25  &   --274  &  8  & A  & WYFFOS      \\
  B245  & 16.56 &   2.99  &    6200  &  1400  & C  & WYFFOS      &    B073  & 15.99 &   2.89  &   --473  &  44  & A  & WYFFOS   \\
  B247  & 17.66 &         &   --532  &    17  & C  & WYFFOS  &    B082  & 15.54 &   4.09  &   --371  &  7  & A  & WYFFOS      \\
  B248  & 17.84 &   1.94  &   --524  &    21  & C  & WYFFOS  &    B083  & 17.09 &   2.48  &   --296  &  19  & A  & WYFFOS     \\
  B253  & 18.01 &   1.76  &   --722  &    35  & C  & DoLoRes     &    B095  & 15.81 &   3.35  &   --238  &  11  & A  & WYFFOS     \\
  B265  & 17.58 &   2.10  &   --496  &    16  & C  & WYFFOS  &    B110  & 15.28 &   3.01  &   --241  &  7  & A  & WYFFOS      \\
  B348  & 16.79 &   2.84  &   --170  &     5  & B  & WYFFOS  &    B117  & 16.34 &   2.46  &   --524  &  25  & A  & WYFFOS     \\
  B362  & 17.61 &   2.09  &    --81  &     4  & A  & WYFFOS  &    B131  & 15.44 &   2.82  &   --337  &  0  & A  & WYFFOS      \\
  B371  & 17.54 &         &   --127  &    18  & B  & WYFFOS  &    B148  & 16.05 &   2.88  &   --261  &  16  & A  & WYFFOS     \\
  B388  & 17.96 &   3.17  &    --50  &     9  & B  & WYFFOS  &    B151  & 14.83 &   3.47  &   --330  &  2  & A  & WYFFOS      \\
  B419(*)& 18.19 &   2.69  &  41600  &   600  & C  & DoLoRes     &    B153  & 16.24 &   3.15  &   --248  &  19  & B  & WYFFOS   \\
  B425(*)& 17.52 &   2.74  &  16900  &   600  & C  & DoLoRes     &    B156  & 16.90 &         &   --400  &  0  & A  & WYFFOS    \\
  B469(*)& 17.58 &   2.60  &  21900  &   600  & C  & WYFFOS      &    B158  & 14.70 &   2.89  &   --190  &  13  & B  & WYFFOS        \\
  B471  & 17.12 &   3.63  &   30200  &   1200 & C  & WYFFOS      &    B163  & 15.04 &   3.36  &   --157  &  23  & A  & WYFFOS   \\
  B473  & 17.46 &   1.51  &      11  &    14  & C  & WYFFOS  &    B174  & 15.47 &   2.98  &   --473  &  16  & A  & WYFFOS     \\
 B020D  & 17.44 &   3.43  &   --526  &    21  & D  & WYFFOS  &    B176  & 16.52 &   2.41  &   --521  &  5  & A  & WYFFOS      \\
 B021D  & 17.50 &   2.59  &       6  &     7  & D  & WYFFOS  &    B178  & 15.03 &   2.41  &   --138  &  6  & A  & WYFFOS      \\
 B022D  & 17.80 &         &   --354  &    24  & D  & WYFFOS  &    B179  & 15.39 &   2.58  &   --151  &  8  & A  & WYFFOS      \\
 B025D  & 17.83 &   3.88  &   --479  &    25  & D  & WYFFOS  &    B180  & 16.02 &   2.62  &   --204  &  13  & A  & WYFFOS     \\
 B027D  & 17.60 &   1.61  &    --50  &     7  & D  & WYFFOS  &    B182  & 15.43 &   2.98  &   --328  &  12  & A  & WYFFOS     \\
 B034D  & 17.50 &   2.97  &   --347  &    25  & D  & DoLoRes     &    B183  & 15.95 &   2.95  &   --179  &  14  & A  & WYFFOS     \\
 B036D  & 17.10 &         &    --54  &     7  & D  & WYFFOS  &    B185  & 15.54 &   2.90  &   --162  &  7  & A  & WYFFOS      \\
 B041D  & 17.90 &   2.65  &   --289  &    11  & D  & WYFFOS  &    B193  & 15.33 &   3.18  &    --65  &  5  & A  & WYFFOS      \\
 B045D  & 18.30 &   2.66  &   --313  &    16  & D  & WYFFOS  &    B201  & 15.90 &   2.47  &   --689  &  17  & A  & WYFFOS     \\
 B046D  & 17.00 &         &   --327  &    24  & D  & WYFFOS  &    B204  & 15.75 &   2.94  &   --351  &  11  & A  & WYFFOS     \\
 B071D  & 17.60 &   2.04  &   --229  &    11  & D  & WYFFOS  &    B206  & 15.06 &   2.57  &   --198  &  13  & A  & WYFFOS     \\
 B073D  & 17.90 &         &    --12  &    13  & D  & WYFFOS  &    B212  & 15.48 &   2.35  &   --413  &  10  & A  & WYFFOS     \\
 B079D  & 17.80 &         &   --394  &    25  & D  & WYFFOS  &    B219  & 16.39 &   2.92  &   --514  &  15  & A  & WYFFOS     \\
 B090D  & 17.20 &   3.63  &    --94  &     8  & D  & WYFFOS  &    B224  & 15.45 &   2.06  &   --162  &  17  & A  & WYFFOS     \\
 B096D  & 17.30 &   3.96  &   --203  &    15  & D  & WYFFOS  &    B225  & 14.15 &   3.08  &   --161  &  13  & A  & WYFFOS     \\
 B109D  & 17.00 &         &     --8  &    15  & D  & WYFFOS  &    B228  & 16.78 &   3.01  &   --400  &  41  & A  & WYFFOS     \\
 B126D  & 18.00 &   2.98  &    --92  &     8  & D  & WYFFOS  &    B229  & 16.47 &   2.22  &   --31   &  5  & A  & WYFFOS      \\
 B147D(*)& 17.96 &         &  38800  &   600  & D  & DoLoRes     &    B230  & 16.05 &   2.23  &   --600  &  5  & A  & WYFFOS      \\
 B148D(*)& 16.31 &   2.20  &  16800  &   300  & D  & BFOSC      &    B232  & 15.67 &   2.40  &   --186  &   9  & A  & WYFFOS      \\
 B158D(*)& 16.50 &   2.69  &  16800  &   300  & D  & DoLoRes     &    B233  & 15.76 &   2.59  &   --72   &  9  & A  & WYFFOS      \\
 B168D(*)& 18.45 &         &  38600  &   600  & D  & DoLoRes    &    B235  & 16.27 &   2.96  &   --92   &  15  & A  & WYFFOS     \\
 B213D  & 17.07 &   1.73  &      19  &     9  & D  & WYFFOS  &    B236  & 17.38 &         &   --411  &  31  & A  & WYFFOS    \\
 B215D  & 16.79 &   2.53  &   --266  &    12  & D  & WYFFOS  &    B238  & 16.42 &   2.73  &   --43   &  14  & A  & WYFFOS     \\
 B217D  & 17.88 &   2.14  &   --136  &     9  & D  & WYFFOS  &    B240  & 15.21 &   2.42  &   --57   &  6  & A  & WYFFOS      \\
 B221D  & 17.77 &   3.32  &    --50  &     8  & D  & WYFFOS  &    B344  & 15.95 &   2.58  &   --240  &  16  & A  & WYFFOS     \\
 B226D  & 17.89 &   3.38  &     --9  &     8  & D  & WYFFOS  &    B347  & 16.50 &   2.37  &   --224  &  24  & A  & WYFFOS     \\
 B237D  & 18.02 &   2.82  &      10  &     8  & D  & WYFFOS  &    B356  & 17.34 &   3.07  &   --179  &  13  & A  & WYFFOS     \\
 B243D  & 18.05 &   2.14  &    --53  &    18  & D  & WYFFOS  &    B366  & 15.99 &   2.01  &   --127  &  21  & A  & WYFFOS     \\
 B250D  & 17.46 &   4.31  &   --442  &    21  & D  & WYFFOS  &    B373  & 15.64 &   3.15  &   --215  &  13  & A  & WYFFOS     \\
 B255D  & 17.92 &         &   --107  &    14  & D  & WYFFOS  &    B377  & 17.14 &   2.68  &   --121  &  32  & B  & WYFFOS     \\
 B260D  & 17.07 &   2.45  &    --93  &     6  & D  & WYFFOS  &    B381  & 15.76 &   2.69  &   --69   &  14  & A  & WYFFOS     \\
 B275D  & 18.11 &   2.26  &    --13  &     6  & D  & WYFFOS  &    B472  & 15.19 &   2.56  &   --101  &  8  & C  & WYFFOS      \\
 NB65  & 16.26 &   2.55  &        8  &     6  & E  & WYFFOS  &    B514  & 16.28 &   2.62  &   --458  &  23  & A  & BFOSC     \\
\hline

\end{tabular}
{\begin{flushleft}{(*): Sources with line-emission spectrum.\\
($\dagger$): original quality-class flag according to Battistini et al.
\cite{bat80,bat82,bat87}.\bigskip\bigskip\bigskip}
\end{flushleft}}
\end{table*}

   \begin{figure}
   \centering
   \includegraphics[width=0.8\hsize]{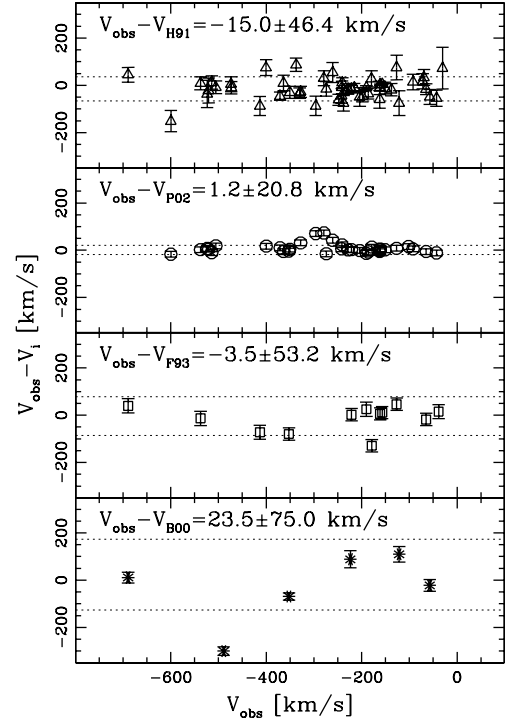}
    \caption{Comparison of radial velocities from the present study ($V_{obs}$)
    with estimates from other authors (H91 = Huchra et al. \cite{bh};
    P02 = Perrett et al. \cite{perr_cat}; F93 = Federici et al. \cite{luciana};
    B00 = Barmby et al. \cite{barm00}). In each panel we report the mean radial velocity
    difference and standard deviations between the two sets under consideration.
    2-$\sigma$ contours around the mean are also reported (dotted lines).
    }
           \label{vr_conf}%
    \end{figure}
%
%
%
   \begin{figure}
   \centering
   \includegraphics[width=0.8\hsize]{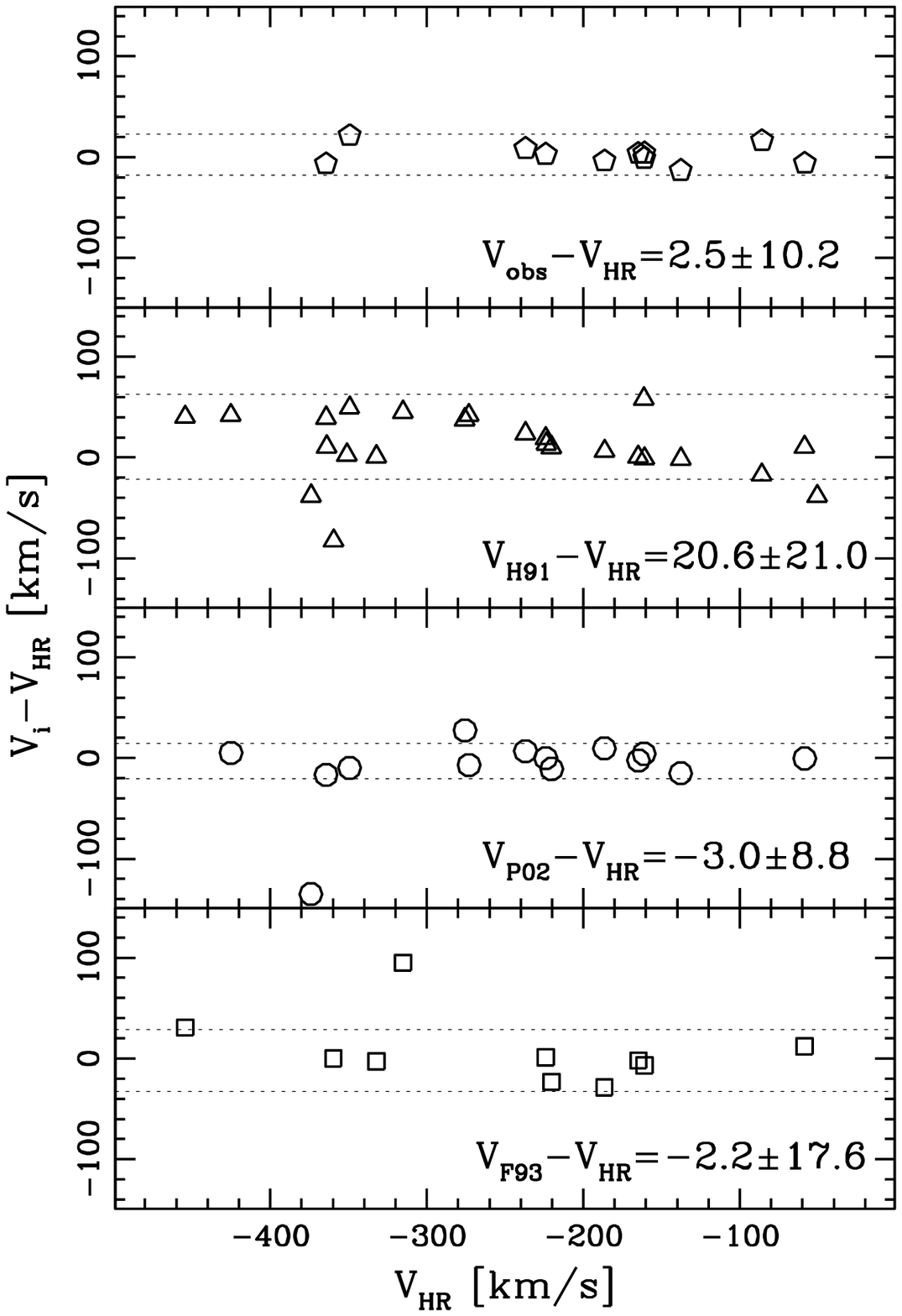}
    \caption{Comparisons of radial velocities from low-resolutions studies with
high-resolution estimates by Peterson (1989) and Dubath \& Grillmair (1997).
The mean differences and standard deviations after 2-$\sigma$ clipping are
indicated in each panel.
    2-$\sigma$ contours around the mean are also reported (dotted lines).
}
           \label{vr_hr}
    \end{figure}
\subsection{Comparisons with previous studies}

Our subsample of 57 already confirmed M~31 GCs in Table~\ref{vr_sur}
allows a thorough comparison with other samples of radial velocities for M~31
clusters available in the literature
[van den Bergh \cite[][ henceforth V69]{syd69};
Huchra et al. \cite{huchra82};
Huchra et al. \cite[][ henceforth H91]{bh}; Peterson \cite{pet};
Dubath \& Grillmair \cite{DG}; Federici et al. \cite[][ F93]{luciana};
Jablonka et al. \cite[][ J98]{jab}; Barmby et al. \cite[][ B00]{barm00};
Perrett et al. \cite[][ P02]{perr_cat}].

This exercise has the twofold purpose of {\it i)} checking the reliability of our measures
and their overall consistency with previous estimates, and {\it ii)} probing the mutual
self consistency of different existing $V_r$ datasets and trying to match each of them
into a common radial-velocity scale.
For this we need to recover the possible systematic offsets in the $V_r$ zero points and provide
a suitable average for those objects with redundant/multiple measurements,
assessing the intrinsic accuracy of the different data sources.

To do that, it is useful to divide the overall database in two categories, i.e.
$V_r$ measurements coming from low-resolution spectroscopy
($\sigma(V_r) =10$-100~km~s$^{-1}$), and those derived from high-resolution
echelle spectra ($\sigma(V_r) \ll 10$~km~s$^{-1}$). In particular, the latter class
includes the works of Dubath \& Grillmair \cite{DG} and  Peterson \cite{pet}, who provided radial
velocities of M~31 clusters with a notably high accuracy ($\sigma(V_r) \lesssim 3$~km~s$^{-1}$) and
excellent agreement for the objects in common. We therefore decided
to merge both samples into a single High Resolution (HR) set of 24 clusters,
referring however to the Dubath \& Grillmair \cite{DG} $V_r$ value for the clusters in common.
The HR set will provide the backbone of a comprehensive velocity scale
joining measures from all the available sources.

In Fig.~\ref{vr_conf} we present a comparisons of our output for the
cluster subsamples in common with the low-resolution observations by
H91, P02, F93 and B00. In all cases the overall agreement is good,
within the combined uncertainties of the two considered datasets.
The comparison with H91 suggests the presence of a small systematic,
also confirmed by the comparison with HR measures. The agreement
with P02 is particularly good. Since the typical accuracy of this
set is very similar to ours, the relatively low dispersion around
the mean ($\sigma=20.8$~km~s$^{-1}$) is a further support to our
estimated $V_r$ uncertainty. The large scatter of the differences
with $V_r$ estimates by B00 is not particularly satisfactory, but
the handful of clusters in common prevents any further conclusion.
Unfortunately, none of the HR cluster is included in the B00 sample.
The comparison of the B00 velocities with the P02 sample (not shown
here) reveals a good general agreement but also several cases of
serious inconsistency with 7 out of 21 clusters in common exceeding
a $\pm 100$~km~s$^{-1}$ difference in the velocity estimates.

\subsection{Consistency checks: towards a single homogeneous list of radial velocities}

The comparisons with the HR set is presented in Fig.~\ref{vr_hr}, where each panel
reports the mean $V_r$ difference and standard deviation for the different samples
after a 2-$\sigma$ clipping iteration.

The agreement of our $V_r$ estimate with the HR ones, for the 12
clusters in common, is excellent. There is no sizable zero-point
difference and again the standard deviation witnesses the good
accuracy of our measures. The same is true for the P02 and the F93
sets, although the latter shows a larger scatter due to a lower
intrinsic accuracy of the data. The match with the H91 sample
confirms, on the contrary, the presence of a systematic, as already
suggested before and noted by H91 themselves. We eventually applied
the zero-point correction of Fig.~\ref{vr_hr} to this dataset, while
we left untouched the P02 and F93 samples, the zero-point
differences being much smaller than the involved statistical
uncertainties. As for the B00 and J98 data, both samples have no
clusters in common with the HR sample and we had to compare them
with other low-resolution observations. No systematic offset is
needed, although one should notice several cases of striking
outliers in each sample.

V69 presented a list of radial velocities for 44 bright M31 GCs and
GCCs. The comparison with the HR velocities for the 19 clusters in
common reveals a small systematic ($V_{V69} - V_{HR} = -15.3 \pm
28.0$~km~s$^{-1}$), after one outlier rejection. Since all of the
V69 targets have been re-observed by several authors in more recent
studies, typically with a better accuracy, we decided to retain the
V69 estimates only as an external check for controversial cases.

Once verified the self-consistency of all the available datasets we merged all
the sources into one single catalog of M~31 cluster radial velocities, tied to the HR set.
This resulted from a weighted average of multiple measures, carefully checked on a
cluster-by-cluster basis.
In general, a weighted mean was iterated after 2-$\sigma$ clipping. When only
two incompatible measures (at $>2 \sigma$) were available, we chose the one with
lowest uncertainty. The overall compatibility among multiple measurements from different
sources is, in general, very good. We found 18 cases of marginal $>2 \sigma$ deviation
of one measure and 9 cases of serious incompatibility, a few of which
already noted by P02. The latter nine cases probably deserve further checks
and we shortly mention them: for B104 we adopted $V_r=-395\pm 10$~km~s$^{-1}$ from B00,
while J98 reports $V_r=+120\pm 42$~km~s$^{-1}$; for B109 we adopted
$V_r=-372\pm 12$~km~s$^{-1}$ from P02 while H91 reports $V_r=-633.6\pm 24$~km~s$^{-1}$;
for B064D\footnote{See Updates \& Revisions {\tt http://www.bo.astro.it/M31/}.} we adopted $V_r=-72\pm 10$~km~s$^{-1}$ from B00 while J98 reports
$V_r=+191.0\pm 62$~km~s$^{-1}$; for B119 we adopted the weighted mean between the
estimates by P02 and J98, $V_r=-310.1\pm 11.1$~km~s$^{-1}$, rejecting
$V_r=-137\pm 10$~km~s$^{-1}$ by B00; for B124 we adopted $V_r=+70\pm 13$~km~s$^{-1}$ from B00,
while J98 reports $V_r=-75\pm 22$~km~s$^{-1}$ (the possibility of a typo should also be
considered in this case); for B301 we adopted the weighted mean between the
estimates by P02 and F93 $V_r=-381.9 \pm 10.9$~km~s$^{-1}$, rejecting $V_r=-30\pm 20$
~km~s$^{-1}$ by B00; for B337 we adopted $V_r=+50 \pm 12$~km~s$^{-1}$ from B00, while F93 reports
$V_r=-232 \pm 26$~km~s$^{-1}$; for B350 we adopted the weighted mean between the
estimates by B00 and H91 $V_r=-467.3\pm 12.7$~km~s$^{-1}$, rejecting
$V_r=-251.7\pm 26$~km~s$^{-1}$ by
F93; finally, for B380 we adopted $V_r=-13\pm 12$~km~s$^{-1}$ from P02 while B00 reports
$V_r=-121 \pm 31$~km~s$^{-1}$. At least for the cases in which only two incompatible
estimates are available (B104, B109, B124, B337, B380, B064D), a third
independent estimate is highly desirable. Also among the 18 cases of
not-so-strong incompatibility there are four cases in which we had to choose
between two estimates based on the accuracy of the single estimates alone
(B008, B047, B144, B314).
For B131 we obtained $V_r=-337 \pm 3.0$~km~s$^{-1}$ while H91 report
$V_r=-444.6 \pm 28$~km~s$^{-1}$; for this case we had to recur to V69 reporting
$V_r=-450$~km~s$^{-1}$, supporting the H91 estimated, that we eventually adopted.

In the following we will use the merged dataset described above
as our source of $V_r$ measures for M~31 GCs.

%
   \begin{figure}
   \centering
   \includegraphics[width=0.8\hsize]{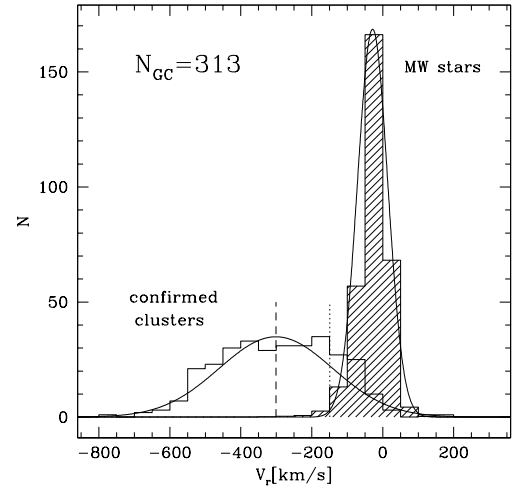}
    \caption{Velocity distribution of confirmed M31 clusters. The adopted
    systemic velocity of M31 is marked by a dashed segment; the dotted segment
    marks the radial velocity threshold beyond which the contamination by
    foreground stars may become a serious issue. A Gauss curve with mean
    $\mu=-301.0$~km~s$^{-1}$ and standard deviation $\sigma=160.0$~km~s$^{-1}$ is also
    superposed. The dashed histogram is the distribution of Galactic stars as
    predicted by the Besan\c{c}on model (Robin et al. \cite{robin}), under the
    assumptions described in Sect.~4. A Gauss curve with mean
    $\mu=-29.0$~km~s$^{-1}$ and standard deviation $\sigma=42.6$~km~s$^{-1}$ is also
    superposed.
     }
         \label{mod}
    \end{figure}
%
%

\section{Classification}

Figure~\ref{mod} displays the radial velocity distribution of confirmed M31 GCs
(empty histogram). The distribution is centered on the galaxy systemic velocity
(i.e.\ $V_s=-301.0$~km~s$^{-1}$, van den Bergh \cite{sydbook}) and has a standard
deviation $\sigma \simeq 160.0$~km~s$^{-1}$. The distribution is significantly
flatter than a Gaussian curve because most of M~31 GCs partake the overall disk
rotation around the galaxy center (see B00 and references therein).
As we mentioned in Sec.~2, just on the basis of the kinematical piece of information,
distant background galaxies can easily be excluded as they have typically
$V_r\gg+300$~km~s$^{-1}$.

To study the contamination by foreground MW stars we obtained a synthetic
sample of Galactic stars over the magnitude range $16.0\le V\le 19.0$, as
spanned by our 76 GCCs, in a field of $3\degr\times 3\degr$ around the position
of M31, from the Besan\c{c}on Galactic Model (Robin et al.\cite{robin}).
Down to this magnitude limit and across the observed field the synthetic
sample predicts more than $10^4$ stars; their radial velocity distribution
is approximately Gaussian, with mean $\langle V_r\rangle = -29.6$~km~s$^{-1}$ and
$\sigma=42.6$~km~s$^{-1}$, in excellent agreement with the sample of confirmed 
stars in the RBC (G04). 

Since the available sample of confirmed
GCs (obviously) has not been obtained from the observation of all the sources
down to $V=19.0$ in a $3\degr\times 3\degr$ field, a direct comparison between
the two samples would be greatly misleading. As a reasonable and conservative
normalization, we assume that the global catalog of 1164 objects
contains as many foreground stars as (presently) confirmed GCs
having $V_r$ estimates, i.e. 313 stars. To limit the
effects of fluctuations due to low-number statistics, we extracted at random
from the whole synthetic catalog 100 samples of 313 stars and we obtained
the $V_r$ distribution for each of them. The shaded histogram is the average of
these 100 distributions.

From Fig.~\ref{mod} it can be safely concluded that GCCs with
$V_r\le-301.0$~km~s$^{-1}$
cannot be MW stars, hence they must be bona fide M31 clusters. According to the
above assumptions the expected number of MW stars with $-301.0< V_r\le -150.0$
km/s is $\le 4$, i.e. $\sim 1$\% of the whole sample. On the other hand, for
$V_r>-150.0$ the contamination by MW stars is likely very significant, preventing a
fully reliable
discrimination between foreground stars and M~31 GCs based on the radial
velocity alone. According to these considerations, we decided to classify as bona
fide M31 GCs all the candidates with $V_r\le -150.0$~km~s$^{-1}$, requiring further
investigation for those with higher $V_r$ (see Sect.~4.1, below).
Note that our approach is more conservative with respect to the majority
of previous studies that, in general, considered as genuine M~31 GCs all the
candidates with velocity within $\sim \pm 3 \sigma$ from the systemic
velocity of M31, in absence of clear evidence of discriminating features in the spectra and/or
HST imaging revealing the stellar nature of the candidate (see, for example,
B00, P02).
On the contrary, Fig.~\ref{mod} strongly suggests that GCCs with
$V_r$ around $\sim -30.0\pm 120$~km~s$^{-1}$ (65 objects in the RBC database,
most of them ``confirmed'' GCs by virtue of their radial velocity alone)
should be very carefully considered as they might likely include a certain fraction
of misclassified MW stars. Clearly, any observation assessing the non-pointlike nature
of these objects would be extremely valuable in this sense (see Sect.~4.1, below).

Using the above described criteria, for our 76 GCCs of Table~\ref{vr_sur} we are left with
\begin{itemize}
\item [(a)] 35 genuine M~31 GCs with $V_r\le -150.0$~km~s$^{-1}$;
\item [(b)] 12 background galaxies, with high recession velocity and/or line-emission spectra;
\item [(c)] 2 M~31 H{\sc ii} regions (never observed before);
\item [(d)] 27 candidates, with $-150.0 < V_r < +100$~km~s$^{-1}$,
possibly compatible both with M~31 GCs {\it and} with MW stars, that
will be further analyzed in the following.\footnote{According to our
constraints on the value of $V_r$, also the controversial object
B341 in Table~\ref{vr_sur} (class $c = 3$ in the original RBC
classification) should eventually be comprised in the {\it
bona-fide} confirmed M~31 GCs supporting the original classification
by P02. The case of B409 deserves, on the contrary, further
discussion (see Sec.~\ref{contro}).}
\end{itemize}

\subsection{Source morphology and foreground star contamination}

To further investigate the nature of the 27 GCCs of item ``d'' above we
acquired deep white-light BFOSC images of each field. In imaging mode,
BFOSC has a pixel scale of $0.58\arcsec$/px and a total field of view of
$13.0\arcmin \times 12.6\arcmin$.
All the observations have been obtained in 2005, during the nights
of Aug 7-9, Sep 2, 11 and 29, Oct 4 and Nov 7.
The exposure times ranged between 3 and 10 minutes, depending on
target brightness and atmospheric conditions. The nights were clear
(but not photometric) with a typical seeing around 1.5-2 arcsec FWHM.

The pointings were accurately chosen in order to include in each frame one (or more)
of the 27 targets and a number of confirmed M~31 GCs. This eventually allowed to extend our
morphological analysis to a supplementary sample of 56 ``confirmed'' GCs plus 3 controversial
targets.\footnote{We also verified {\it a posteriori} that, out of these 59 objects,
13 further entries from Table~\ref{vr_sur} were serendipitously imaged. They are B042,
B100, B193, B224, B228, B229, B232, B240, B265, B344, B347, B366, B045D (see Table~\ref{gc}).}

Images have been bias and flat-field corrected with standard reduction
procedures. Relative photometry, FWHM and morphological parameters of each source
in the frame - down to a 5~$\sigma$ threshold over sky noise- were obtained with Sextractor
(Bertin \& Arnouts \cite{sex}).
Only non-saturated and isolated sources were retained in the final catalogs
(Sextractor quality flag ``0'').

\begin{table*}[!ht]
\scriptsize
\caption{Morphological analysis for previously confirmed M~31 GCs.}             
\label{gc}      
\centering

\begin{tabular}{l r@{$~~\pm$} r r c r  | l r@{$~~\pm$} r r c r }
\hline
        &   \multicolumn{2}{c}{}   &    &     &  &    &  \multicolumn{2}{c}{} &  &   &   \\
Name   & Vr    & ~~$\sigma$($V_r$)& R & F-A flag & c$^{(\dagger)}$  & Name  & Vr    & ~~$\sigma$($V_r$)& R & F-A flag & c$^{(\dagger)}$  \\    
       &  \multicolumn{2}{c}{~~~[km~s$^{-1}$]} &  & &   &   &  \multicolumn{2}{c}{~~~[km~s$^{-1}$]} &  & & \\    
        &   \multicolumn{2}{c}{}   &    &     &  &    &  \multicolumn{2}{c}{} &  &   &   \\
\hline                                    
 B213  &--545  & 11 &1.15 &   E   &  1&  B347  &--251  & 20    &1.14           &   E   &  1\\
 B265  &--496  & 16 &1.10 &   E   &  1&  B048  &--251  & 12   &1.22    &   E   &  1\\
 B209  &--460  & 11 &1.18 &   E   &  1&  B220  &--247  & 12   &1.75    &   E   &  1\\
 B015D &--445  & 12 &1.31 &   E   &  1&  B167  &--231  & 10    &1.15           &   E   &  1\\
 B093  &--447  & 12 &1.25 &   E   &  1&  B075  &--212  & 12    &1.38           &   E   &  1\\
 V031  &--433  & 12 &2.33 &   E   &  1&  B203  &--199  & 12   &1.99    &   E   &  1\\
 B161  &--413  & 12 &1.20 &   E   &  1&  B154  &--199  & 33    &1.40           &   E   &  1\\
 B221  &--406  & 12 &1.24 &   E   &  1&  B188  &--184  & 12    &1.28           &   E   &  1\\
 B021  &--403  & 12 &1.29 &   E   &  1&  B232  &--182  &  7    &1.09           &   E   &  1\\
 B228  &--457  & 26 &1.43 &   E   &  1&  B224  &--161  &  2    &2.21           &   E   &  1\\
 B100  &--376  & 25 &1.45 &   E   &  1&  B200  &--153  & 12   &2.12    &   E   &  1\\
 B467  &--342  & 12 &1.26 &   E   &  1&  B184  &--152  & 12    &1.39           &   E   &  1\\
 B043D &--344  & 12 &1.51 &   E   &  1&  B367  &--152  & 12   &1.07    &   E   &  1\\
 B037  &--338  & 12 &1.30 &   E   &  1& \multicolumn{6}{c}{\hrulefill} \\
 B042  &--338  & 10 &1.14 &   PS  &  1&  B103D &--148  & 12   &1.18       &   E   &  1\\
 B401  &--333  & 23 &1.26 &   E   &  1&  B240D &--148  & 12   &5.83       &   E   &  1\\
 B059  &--332  & 12 &1.24 &   E   &  1&  B366  &--141  & 10   &1.00     &   E &  1(3) \\
 B391  &--325  & 12 &3.73 &   E   &  1&  B272  &--120  & 12   &1.53       &   E   &  1\\
 B045D &--313  & 16 &1.22 &   E   &  1&  B355  &--114  & 12  &0.99 &   PS  &  1(6)\\
 B024  &--310  & 34 &1.09 &   E   &  1&  B198  &--105  & 12  &1.20    &   E   &  1\\
 B222  &--303  & 10 &1.36 &   E   &  1&  B046  & --98  & 49   &1.13    &   E   &  1\\
 B382  &--302  & 12 &1.12 &   E   &  1&  B216  & --93  & 10    &1.36 &   E   &  1(7)\\
 B164  &--294  & 12 &1.20 &   E   &  1&  B072  & --89  & 12   &2.52       &   E   &  1\\
 B091  &--290  & 12 &1.31 &   PS  &  1&  B190  & --86  & 12   &1.17       &   E   &  1\\
 B047  &--291  & 12 &1.19 &   E   &  1&  B193  & --59  &  2   &1.07       &   E   &  1\\
 B354  &--283  & 26 &1.14 &   E   &  1&  B240  & --56  &  5   &1.23       &   E   &  1\\
 B210  &--265  & 12 &1.18 &   E   &  1&  B229  & --31  &  5   &1.36       &   E   &  1\\
 B214  &--258  & 12 &1.12 &   E   &  1&  B197  &  --9  & 12   &1.27 &   E   &  1\\
 B344  &--252  & 13 &1.34 &   E   &  1& \multicolumn{6}{c}{} \\
\hline
\end{tabular}
{\begin{flushleft}{$(\dagger)$ New classification flag, according to the RBC notation
(see also Sect.~1). The newly determined class of the three object for which
 we modify previous classification is reported in parentheses.}
\end{flushleft}}
\end{table*}

In spite of the limited spatial resolution of the images, a fair
assessment of the GCC morphology was made possible by a purely differential approach
relying on the comparison with accurate point-spread function (PSF) estimates
and on the study of the apparent isophotal radius of the detected objects.
The relatively wide field of view of BFOSC allowed in fact a simultaneous high-S/N imagery
of the target together with hundreds of field stars, and at least a couple of confirmed M~31 GCs.
Hence, the nature of the considered candidate was established by direct
comparison with surrounding stars and extended sources {\em on the same frame}.

\begin{figure}
   \centering
   \includegraphics[width=0.8\hsize]{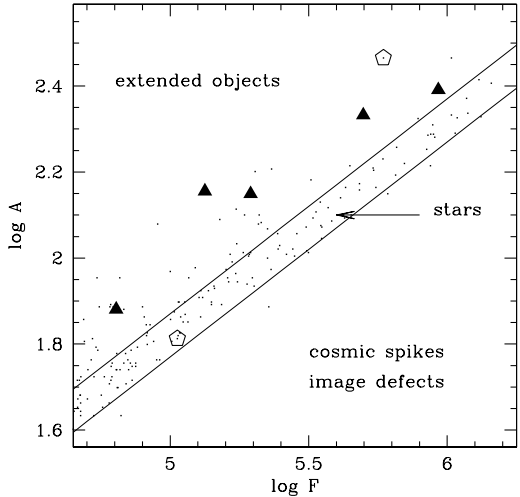}
    \caption{  F-A diagram for one of the images we analyzed (as an example).
    All the detected sources are plotted as small points. Filled triangles
    correspond to already confirmed clusters, empty pentagons to the candidate
    clusters under consideration. The continuous lines enclose the locus of
    points sources.
     Extended objects are expected to populate the upper left
    corner above
    the strip, while cosmic spikes and CCD cool/hot pixels must be confined to
    the lower right corner of the plot.
}
           \label{AF}
    \end{figure}

In particular, the apparent target size was probed {\it a)} in terms of its relative excess
compared to the local PSF (namely, by defining a ratio parameter
$R = {\rm [target~FWHM]}/ {\rm [PSF~FWHM]}$) and
{\it b)} in terms of its departure from the isophotal flux ($F$) vs.\ isophotal area ($A$)
empirical relationship for stars to be suitably set in each individual frame according to the overall
image quality.\footnote{Note that the limiting
isophote is set at a $S/N = 5$ ratio per pixel element on each frame. This corresponds, in general,
to a different surface brightness magnitude depending on image quality and sky conditions.}
Both $F$ and $A$ are natural outputs of Sextractor (respectively {\sf FLUX\_ISO} and
{\sf ISOAREA\_IMAGE} parameters), and one could verify empirically that essentially all stars
roughly obey a $A \propto F^{1/2}$ relationship being enclosed within a 0.1 dex wide
strip in the $\log F$ vs.\ $\log A$ plane, corresponding to a variation of the measured FWHM
of $\pm 5$\% (see Fig.~\ref{AF}).
Of course, hot pixels and cosmic rays appear as off-strip ``sub-point-like'' objects, while any source
significantly more extended than the stellar PSF stands out in the F-A plot for its larger apparent
size for a given magnitude level. To some extent, our approach recalls the classical diagnostic
tools used, for instance, for high-redshift galaxy recognition (e.g.\ Kron \cite{kron},
Koo et al. \cite{koo}, Molinari, Buzzoni \& Chincarini \cite{molinari}).

Figure~\ref{AF} is an illustrative example for a field matching two GCCs (open pentagons) and five
already confirmed clusters (solid triangles). While all the five confirmed GCs are univocally
identified as extended objects in the F-A plane, one of the candidates appears as an extended
source, while the other one is compatible with a stellar point source. Accordingly, from the inspection of
the individual F-A diagrams, we assigned to all the 27 candidates under consideration the
flag E (extended) or PS (point source). In the following we will refer to this flag as the
F-A flag or, for brevity, the flag.

The R parameter and the F-A plane provide two nicely complementary tools to
judge the candidate extension since R is mainly sensitive to the core of
the target image, while A, as the area within the outermost isophote, is more
sensitive to the wings of the image. This is very well suited to study GC candidates since
profiles of globular clusters may significantly vary, depending on their
concentration parameter C (King \cite{king}). 
For instance, high concentration clusters
may have such compact cores that are essentially indistinguishable from point
sources but a significant difference with respect to genuine point sources can
be detected looking at the wings of the image, sampling the faint halo of the
cluster (see, Buonanno et al. \cite{buon}).

It may be useful to have an idea of the sensitivity of the adopted technique,
i.e. what is the size of the most compact M31 GC that can be recognized as an
extended object by our diagnostics. To this aim we performed a series of tests
using the half-light radii ($r_h$) of globular clusters as the 
characteristic scale that is more appropriate in this context. 
According to Djorgovski \cite{Djo}, half-light radii of Galactic GCs range from
0.8 pc to 20 pc, and more than 75\% of his sample (118 GC with estimates of the
half-light radius) have $r_h>2.0$ pc. We translated these linear radii into
angular radii at the distance of M31 ($D=783$ kpc, see Sect.~5.1, below) and we
convolved them with the Half Width at Half Maximum (HWHM) of the PSF
of our images. Finally, we found the minimum $r_h$ that
provides a convolved profile larger than that of a point source by a factor
$\ge 1.05$. 
It turns out that we are able to pick up clusters having $r_h$ as small as
1.5 pc (corresponding to $0.4\arcsec$, at the distance of M31), 
if the seeing is $2.0\arcsec$ FWHM, and we are sensitive to $r_h>2.0$ pc if 
the seeing is $3.0\arcsec$ FWHM. Note that the radii of the
isophotes adopted to compute {\sf ISOAREA\_IMAGE} are typically a factor 2 larger
than half-light radii, hence the effective sensitivity of the F-A diagram should
be significantly better than these figures. This is confirmed by a direct
comparison that can be made on three M31 clusters included in the present 
programme that have an estimate of $r_h$ from HST data (Barmby \& Huchra
\cite{barm01}), namely B167 ($r_h=0.33\arcsec$), B232 ($r_h=0.66\arcsec$), and
B240 ($r_h=0.80\arcsec$). All of them are clearly identified as extended objects
in the F-A diagram and their R parameters are 1.15, 1.09 and 1.23, respectively
(see Table 2). Therefore, we conclude that the image analysis technique adopted
here is able to recognize the extended nature of the large majority of M31
clusters even with low-resolution images, provided that the target image is of
high signal-to-noise so that the image profile is well constrained.

\subsubsection{Checking previously confirmed ``genuine'' clusters with $V_r\le -150.0$}

Table~\ref{gc} reports the results of the image analysis described
above for the M~31 clusters previously confirmed by other authors
that were included in our BFOSC imaging survey. The first 42 entries
in the table concern clusters that have $V_r\le -150.0$~km~s$^{-1}$,
and can be unambiguously identified as genuine GCs on the basis of
their radial velocity alone, according to our previous arguments. We
note that all of them have $R\ge 1.07$, and all except three have
$R\ge 1.10$. Only two clusters obtain the PS flag based on their
position in the F-A diagram, but both of them have quite large R,
1.14 and 1.31 for B042 and B091, respectively. Hence our morphologic
criteria correctly recognize all the genuine M~31 GCs considered
here as extended objects.

Also based on these results, we can complete our classification scheme and devise the following
supplementary criteria for GCCs with  $V_r> -150.0$~km~s$^{-1}$:
\begin{itemize}

\item [(a)] a candidate is classified as a ``bona fide M31 cluster'' ($c=1$ in Tables 2, 3, and 4)
if $R\ge 1.10$, independently from the assigned flag, or if $R> 1.05$ {\em and}
F-A flag ``E''.

\item [(b)] it is classified as a ``bona fide star''
($c=6$) if $R\le 1.05$ {\em and} F-A flag ``PS'' or if R$\le 1.0$.

\item [(c)] it is considered as an ``uncertain object'' ($c = 3$) if
$1.0<R\le 1.05$ {\em and} flag ``E'', or $1.05< R < 1.10$ {\em and}
flag ``PS''. These non-decidable cases can be resolved only with
further data and/or analysis.

\end{itemize}

\begin{figure*}
   \centering
   \includegraphics[width=0.6\hsize]{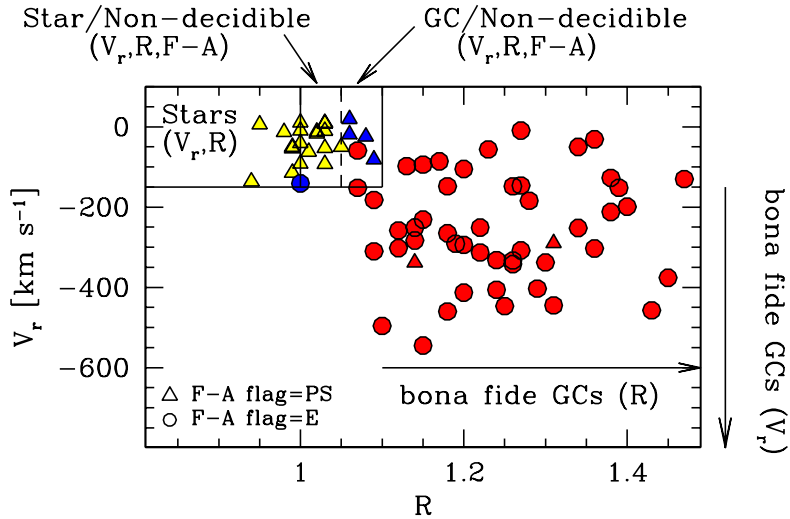}
    \caption{  Summary of the adopted classification criteria illustrated in the
    R vs. $V_r$ diagram, where different symbols are used according to the value
    of the F-A flag of the targets. Circles represent extended
objects (F-A flag = E) while triangles are point-like sources (F-A
flag = PS). {\em Bona fide} stars (c=6) are colored light-grey, {\em
bona fide} GCs (c=1) are plotted in darker grey, while the sources
that cannot be firmly classified within our scheme (c=3) are colored
in the darkest grey. The box in the upper-left corner encloses stars
and candidates whose nature requires the use of all the three
parameters to be classified ($V_r$, R, F-A flag). The regions of the
plot corresponding to different classifications are labeled. Note
that most of the  clusters with R$\ge 1.5$ (Tab.~2 and 3) do not appear in
this plot for the sake of graphical clarity.
}
           \label{scheme}
    \end{figure*}

The overall classification picture for the whole GCC sample of Table 2 and the following ones
is summarized in Fig.~\ref{scheme}.

%

\begin{table}
\caption{Morphological analysis for the 27 M31 globular clusters candidates
of Table~\ref{vr_sur} with $V_r>-150.0$~km~s$^{-1}$.
}             
\label{star}      
\scriptsize
\begin{tabular}{l r@{$~~\pm$} r c c c c r  }
\hline
Name   & Vr    & ~~$\sigma$($V_r$) & class$^{(a)}$ & R$^{(b)}$ & F-A flag$^{(c)}$ &c$^{(d)}$\\    
       &   \multicolumn{2}{c}{~~~[km~s$^{-1}$]}   &       &     &          & \\    
\hline                      
  B189 &--148 &  8 &  A &1.26 &  E   &1\\
  B162 &--146 &  8 &  A &1.27 &  E   &1\\
 B217D &--136 &  9 &  D &0.94 &  PS  &6\\
  B187 &--130 & 10 &  A &1.47 &  E   &1\\
  B371 &--127 & 18 &  B &1.38 &  E   &1\\
 B255D &--107 & 14 &  D &2.5  &  E   &1 \\
 B126D & --92 & 18 &  D &1.00 &  PS   &6\\
 B090D & --94 &  8 &  D &1.15 &  E   &1\\
 B260D & --93 &  6 &  D &1.03 &  PS   &6 \\
  B362 & --81 &  4 &  A &1.09 & PS   &3\\
 B330D & --62 & 18 &  D &1.01 & PS   &6 \\
 B036D & --54 &  7 &  D &0.99 &  PS   &6\\
 B243D & --53 & 18 &  D &1.03 & PS   &6 \\
  B388 & --50 &  9 &  B &1.34 &  E   &1\\
 B027D & --50 &  7 &  D &0.99 &  PS   &6\\
 B221D & --50 &  8 &  D &1.05 & PS   &6\\
  H126 & --19 & 10 &  / &1.06 & PS   &3\\
  NB70 & --17 & 11 &  E &1.02 & PS   &6 \\
 B275D & --13 &  6 &  D &0.98 &  PS   &6 \\
 B073D & --12 & 13 &  D &1.02 & PS   &6\\
 B226D &  --9 &  8 &  D &1.00 &   PS   &6\\
 B109D & --10 & 15 &  D &1.03 &   PS   &6\\
 B021D &    6 &  7 &  D &0.95 &   PS   &6\\
  NB65 &    8 &  6 &  E &1.03 & PS   &6\\
 B237D &   10 &  8 &  D &1.03 & PS   &6\\
  B473 &   11 &  7 &  E &1.00 &   PS   &6\\
 B213D &   19 &  9 &  D &1.06 & PS   &3\\
\hline
\end{tabular}
{\begin{flushleft}{
(a) Quality flag according to the Battistini et al. \cite{bat80,bat82,bat87,bat93} classification scale;\\
(b) Target apparent size, relative to the PSF (namely $R = {\rm [target~FWHM]}/ {\rm [PSF~FWHM]}$;\\
(c) E(xtended) or P(oint) S(ource) classification obtained from the F-A diagram;\\
(d) RBC classification flag as in Tab.~\ref{gc} (see Sect.~1.).
}
\end{flushleft}}
\end{table}

\begin{table}
\caption{Candidates with controversial classifications}             
\label{gc2}      
\centering

\begin{tabular}{l r@{$~~\pm$} r c c c c r }
\hline
Name   & Vr    & ~~$\sigma$($V_r$) & R         & F-A flag & c\\    
       &   \multicolumn{2}{c}{~~~[km~s$^{-1}$]}  &         &      &  \\
\hline                                            
 B055  & --308 &  8  &1.27  & E  & 1\\
 B121  & --24  & 33  &1.08  & PS & 3\\
 B409  & --40  & 10  &1.00  & PS & 6\\
\hline
\end{tabular}
\end{table}

\subsubsection{Image analysis for previously ``confirmed'' clusters with $V_r> -150.0$~km~s$^{-1}$}

The last 14 entries of Tab.~\ref{gc} concern clusters previously confirmed by
other authors, that have $V_r> -150.0$~km~s$^{-1}$, i.e. a range of velocity that
may suffer from contamination by MW stars. As noted in Sect.~4., those that have
been confirmed only by virtue of their radial velocity may deserve further
analysis to obtain a firm classification. According to the above criteria, 12
of these candidates are classified as genuine M31 GC,
one (B366) is classified as ``uncertain'' and one (B355)
is classified as ``bona fide star''.
It is interesting to note that B355 was classified as a confirmed GC by Perrett et.
al. \cite{perr_cat} relying only on its radial velocity.
Finally, B216 makes a case on its own that will be discussed in detail in
Sect.~4.3.

\subsubsection{Image analysis for candidates with $V_r> -150.0$~km~s$^{-1}$}

The results of the image analysis for the 27 GCCs from our sample of
Table~\ref{vr_sur} with $V_r> -150$~km~s$^{-1}$ are reported in
Table~\ref{star}. According to the above criteria, 7 of them are
classified as genuine GCs (namely B162, B187, B189, B371, B388,
B090D, and B255D), 17 are flagged as stars and 3 remain uncertain.
We note that all the newly confirmed clusters are very clearly
recognized as extended: all of them have flag=E and R ranges between
1.15 and 2.5.  Note also that 5 of the 6 candidates in the range
$-150 < V_r < -100$~km~s$^{-1}$ turned out to be genuine GCs, while
for the 21 targets with $V_r> -100$~km~s$^{-1}$ we only detect 2
clusters and 16 {\it bona fide} stars.  This further confirms how
severely star contamination can affect in the range of $V_r$ around
$\sim -30\pm 130$~km~s$^{-1}$.

\subsubsection{A few controversial objects\label{contro}}

The three entries of Table~\ref{gc2} concern candidates that obtained
different classifications from different authors. B055 was classified GC by P02,
based on its radial velocity, while B00 classified it as a star. Our image
analysis indicates unambiguously that B055 is an extended object, thus
confirming the classification by P02. B121 was classified GC by H91 and ``star''
by B00 and our analysis does not help to resolve the controversy.

Object B409 was classified as a GC by F93, based on its radial
velocity alone, and it was classified as a background galaxy by
Racine \cite{racine91}, based on ground-based, high resolution
imaging. Our morphological analysis of this object, and its low
radial velocity, independently confirmed from Table~\ref{vr_sur}
data, agree for a decidedly clean point-source appearance; according
to the adopted criteria, we therefore classify it as a foreground
star.


   \begin{figure*}
   \centering
   \includegraphics[width=0.6\hsize]{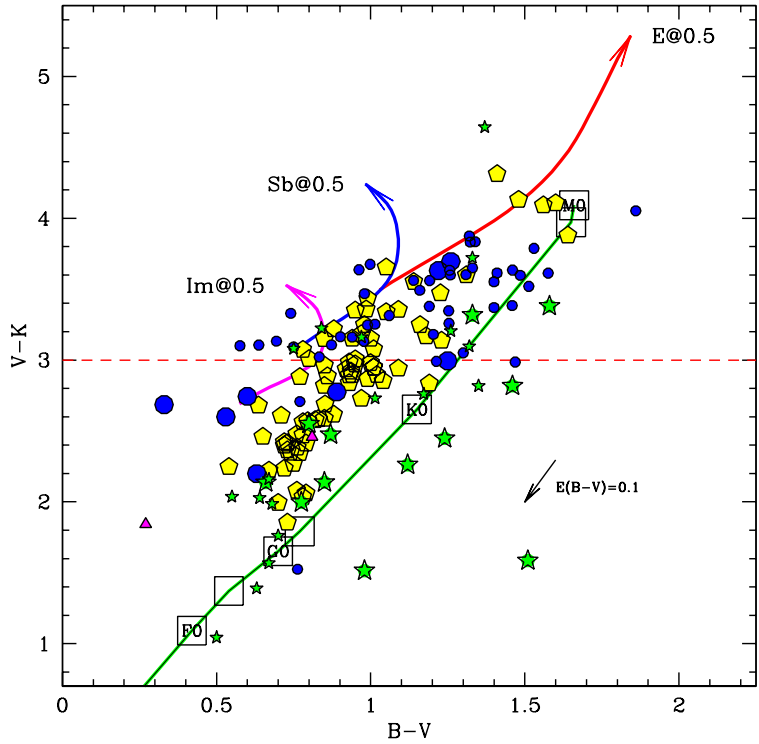}
    \caption{ Apparent $V-K$ vs.\ $B-V$ color distribution for the 45 targets of Table~\ref{vr_sur}
    with available $B,V,K$ photometry (big markers on the plot) compared to the corresponding
    RBC classification groups (small markers). Big pentagons are the {\it bona-fide}
    ($c = 1$ RBC flag) M~31 globular clusters of this study, solid dots are background galaxies
    ($c = 4$), solid triangles mark M~31 H{\sc ii} regions ($c = 5$), and star markers
    are MW stellar interlopers ($c = 6$) and M~31 asterisms ($c = 7$).
    The arrow indicates the reddening vector corresponding to
    $E(B-V) = 0.1$. The main stellar locus for stars of different
    spectral type (F0,F5, G0, G5 dwarfs and K0, K5, and M0 giants from
    Johnson \cite{john} is superposed to the data, as well as the  expected
      apparent colors vs. redshift for the Buzzoni \cite{b05} template
      galaxy models for elliptical, spiral (type Sb) and irregular (type Im)
      Hubble types. Galaxy colors are tracked from $z = 0$ to 0.5 (as labeled
      on the plot). The stellar locus and the galaxy models {\it have been reddened}
      assuming $E(B-V)=0.11$.
      Note that the main background contaminants to M~31 GCCs (i.e.\ ellipticals
      and nucleated spirals within $z \lesssim 0.2$) are always redder than $V-K = 3.0$ while
      only star-forming spirals and irregulars (as well as MW stars) become the prevailing
      contaminants at bluer colors.
}
           \label{cm}%
    \end{figure*}

\subsection{Background galaxy contamination}

Since most of the known GCCs have been historically selected because of their fuzzy and/or
extended look on photographic plates, it is quite natural that the major contamination
source is represented by background galaxies, in particular those of spheroidal morphology.
Any eye-detection survey is in fact reasonably safe with
respect to low-redshift grand-design spirals, that can usually be confidently picked up, at least
in good-quality images.

Spectrophotometric information can help to segregate with some
confidence nucleated galaxies and M~31 GCs in the color domain. In
G04 we noted, for instance, that the large majority of spurious
GCCs, eventually identified as background galaxies, are redder than
$(V-K) > 3.0$, so that by restraining target selection to bluer
objects one should in principle maximize the detection of genuine
clusters. This guess is supported on a more physical basis looking
at the expected photometric evolution of early- and late-type galaxy
models by Buzzoni \cite{b05}, as shown in Fig.~\ref{cm}. On the
other hand, the contribution of blue galaxy contaminants (closer in
color to the bulk of low-metallicity M~31 GCs) may increase as far
as our GC search extends to fainter and more blurred targets
including poorer candidates in the Battistini et al. image-quality
classification.

According to the updated RBC sample, the relative source partition of
{\it bona fide} [galaxies: GCs: stars] among the entries with available $V$ and $K$ photometry
and $(V-K) \ge 3.0$ is found to be [47: 102: 10], while the corresponding frequency blueward of
the $(V-K)$ threshold is [11: 202: 27] .
Taken at their face value, these numbers suggest that about 2/3 of the whole $(V-K) \ge 3.0$
GCCs might eventually confirm to be genuine clusters, while our performance should
raise to a nearly 85\% for $(V-K) < 3.0$ candidates (see G04).

On the other hand, one has to admit that these figures are at odds with the empirical evidence
from the present analysis,
as for the whole sample of 71 GCCs of Table~\ref{vr_sur} with firm classification
we find [galaxies: GCs: stars]~=~[12: 42: 17], with a remaining fraction of 5 H{\sc ii} regions
and unclassified objects (again, see Fig.~\ref{cm} for a summary). Thus, about one
in two of our targets eventually revealed to be a genuine globular, in spite of the fact
that over 70\% of our GCC sample is bluer than $(V-K) \ge 3.0$, and an {\it a priori} distribution
should be expected such as [galaxies: GCs: stars]~=~[9: 57: 7] and 3 unclassified objects.

\begin{figure}
   \centering
   \includegraphics[width=0.85\hsize]{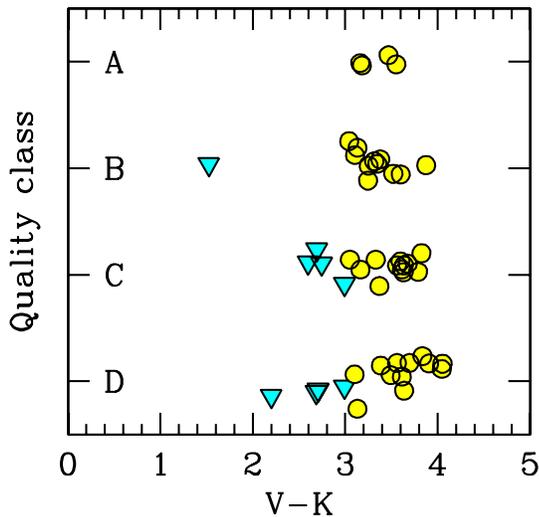}
    \caption{The $V-K$ color distribution for confirmed background galaxies ($c = 4$ flag) in the
    whole RBC sample vs.\ original quality class according to the Battistini et al. \cite{bat80,bat82,bat87,bat93}
    scheme. Blue ($V-K < 3.0$) late-type galaxies (solid triangles) become an increasing contaminant
    source among class C--D cluster candidates. See text for a discussion.}
           \label{bias_sp}%
    \end{figure}

These figures lead us to the following conclusions:
\begin{itemize}
\item [(a)] As far as deeper M~31 GCC surveys are carried out (and naturally include
poorer and more ``blurred'' targets), the galaxy contamination becomes
marginally more important, with an increasing contribution of blue spirals and star-forming systems.
This is especially evident when we plot the galaxy $(V-K)$ distribution along the
different Battistini et al. image-quality classification, as in Fig.~\ref{bias_sp}.
Fortunately, this bias can in principle be fully overcome as any spectroscopic identification
would easily discriminate these spurious emission-line objects (note, for instance, that
9 out of 12 recognized galaxies in our Table~\ref{vr_sur} sample display some
Balmer and/or [O{\sc iii}] emission).

\item [(b)] Between galaxy and star contaminants, the latter become increasingly
important as far as class C-D-E Battistini et al. candidates are surveyed (we find
17 stellar interlopers vs.\ 7 expected cases).
\end{itemize}


\subsection{Contamination by asterisms/associations}

Radial velocities and image analysis are very effective tools to
find out the most abundant sources of contamination, i.e.\
background galaxies, foreground stars and H{\sc ii} regions.
However, they can be mocked by a further and much subtler kind of
contaminants. Small stellar associations, very young M~31 open
clusters (whose integrated luminosity is dominated by a few massive
stars), and even perspective asterisms due to chance alignment of
M~31 stars (and/or one or a few stars embedded into, or projected
onto, a nebula) can easily mimics a typical faint GCC, when observed
in low spatial-resolution images. Clearly, this class of
contaminants is expected to affect faint and blue GCCs, especially
those projected onto the disc of M~31. Cohen, Matthew \& Cameron
\cite{coh} recently provided direct evidence of at least four such
special cases among spectroscopically confirmed clusters, using
exquisite spatial-resolution images obtained with adaptive optics.
To account for these ``new'' kind of contaminants, we introduce a
new classification type, class $c = 7$, corresponding to
asterisms/associations of M31 stars.

In spite of the possible occurrence of these ``fake'' stellar aggregates,
whose real occurrence needs however to be more firmly assessed on a statistical basis,
it is clear that, {\it in any case}, the final word on the real nature of M~31
GCCs must come from the physical resolution of the composing stars, through very
high-resolution imaging (i.e., at HST or with adaptive-optic ground telescopes).

Among the clusters revealed to be asterisms by Cohen et al. \cite{coh}, there
is one, B216, we classify as a genuine GC in Table~\ref{gc}, based on its velocity and
apparent morphology. While the other asterisms in the Cohen et al. list cannot be
independently assessed from our low-resolution imagery, both the BFOSC and the
DSS frames for B216 appear hardly compatible with the nearly empty field
imaged by Cohen et al.. However, we carefully checked
the location of this field (with the precious collaboration of J.\ Cohen) and
the listed target coordinates, and we must conclude that B216
is indeed not a real cluster.

\subsection{Independently checked candidates}

As a duty cycle operation to maintain an updated release for the
RBC, we periodically search the HST archive for intentional or
serendipitous images of M~31 GCCs that can potentially reveal the
true nature of the objects. A systematic survey of the available
material is ongoing. Here we report only a few cases in which a
clear and indisputable confirmation can be achieved from a thorough
inspection of the images, i.e.\ objects partially resolved into
stars or obvious foreground stars or asterisms.

   \begin{figure}
   \centering
   \includegraphics[width=0.7\hsize]{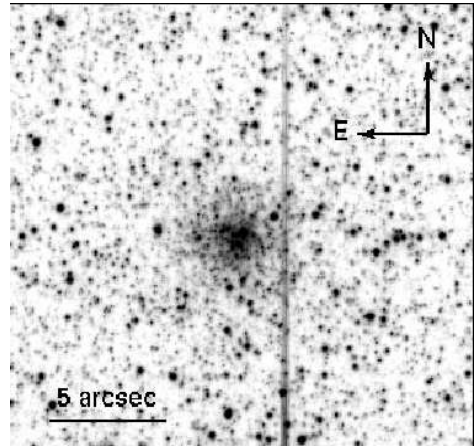}
      \caption{The newly identified cluster B515 from F606W
      HST/ACS imagery. The scale and the orientation of the image are
      reported.
               }
      \label{b515}
   \end{figure}
%

   \begin{figure*}
   \centering
   \includegraphics[width=0.6\hsize]{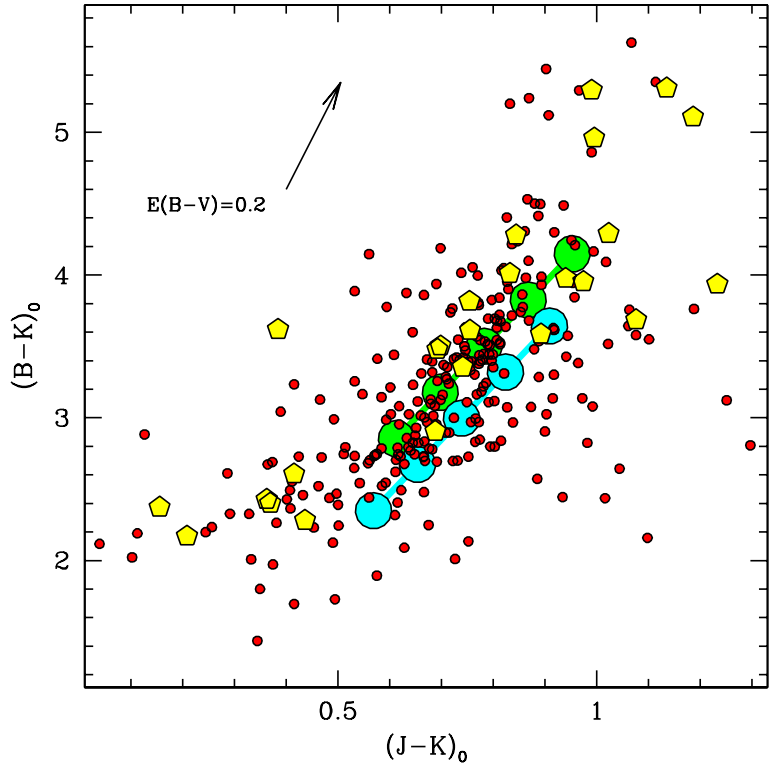}
      \caption{Distribution of confirmed M31 clusters in the reddening-corrected
      $J-K$ vs.\ $B-K$ plane. $E(B-V) = 0.11$ is assumed for all M31 clusters.
      Small filled circles are previously confirmed clusters, pentagon markers are the
      newly confirmed clusters studied in the present paper.
       SSP model sequences by Buzzoni \cite{buz},
      for fixed ages (i.e. 15 Gyr for the upper sequence and 2 Gyr for the lower sequence)
      and varying metallicity (from [Fe/H] = --2 to a solar value, at equal steps of 0.5~dex,
      see big solid dots in the sense of increasing $J-K$ color along the sequences) are
      superposed.
      A Salpeter IMF slope and a red horizontal branch morphology is assumed in the models.
}
      \label{jkbk}
   \end{figure*}

\begin{itemize}

\item A new, clearly resolved, cluster has been identified by L.F. in deep
F555W and F814W ACS/WFC images (see Fig. \ref{b515}).
The cluster has no counterpart
in the RBC and is located at $\alpha_{2000}= 00^h 42^m 28.05^s$,
$\delta_{2000}=41\degr 33\arcmin 24.5\arcmin\arcmin$. According to the
nomenclature adopted in G04 and G05 we christen the newly found cluster Bologna
515 (B515).

\item the candidate B056D is clearly recognized as a genuine cluster on several
deep ($t_{exp}$ up to 2370 s) ACS/WFC images taken in different filters,
by different teams.

\item NB83, classified as genuine cluster by B00 based on its radial velocity
($V_r=-150 \pm 14.0$~km~s$^{-1}$) is clearly recognized as a star in deep F555W and
F814W images taken with the WFPC2. Note that the radial velocity is in the range
where contamination by MW stars can occur (see Sect.~4.1).

\item B102, classified as genuine cluster by H91 and P02 based on its radial
velocity ($V_r=-235.4 \pm 11.7$~km~s$^{-1}$) is recognized as an asterism formed by two
stars superposed to a nebulosity in deep ACS/WFC images taken in different
filters.

\item the candidate NB92 is recognized as a bright (likely foreground) star in
deep ACS/WFC and in shallow WFPC2 images taken in various filters.

\item the candidate B162, that we classified as a genuine cluster based on its
radial velocity ($V_r=-146 \pm 8$~km~s$^{-1}$) and on its extendedness (see
Tab.~3), is clearly recognized as a genuine cluster also on several
deep ($t_{exp}$ up to 2370 s) ACS/WFC images taken in different filters,
by different teams.

\item the candidate G137, that we recognized as an HII region from its spectrum,
appears as a
bright point-source surrounded by an asymmetric nebula on deep ACS/WFC images,
thus confirming our spectroscopic classification.

\item B118\footnote{See Updates \& Revisions {\tt http://www.bo.astro.it/M31/}.}, NB99, NB100, NB106, classified as stars in the RBC are recognized as
stars also in deep ACS/WFC images, thus confirming the existing classification
on a much firmer basis.

\end{itemize}

In summary, we identified a new cluster (B515), one candidate is
recognized as genuine clusters (B056D), two objects previously
believed to be genuine clusters has been re-classified as foreground
star and asterism (NB83 and B102, respectively), one candidate has
been firmly classified as a star (NB92), and the existing
classification of seven other objects (B118, B162, B137, NB99, NB100
and NB106) has been fully confirmed.

\section{An updated sample of confirmed M31 GCs}

As a result of our spectroscopic and imaging survey of Table~\ref{vr_sur} candidates,
we have provided 42 newly
confirmed {\it bona-fide} M~31 clusters, while a total of 34 GCCs should in fact be
comprised among background galaxies, foreground stars or H{\sc ii} regions.
Our study increases the total number of confirmed M~31 GCCs from 337 to 368 and the number
of confirmed GCs having a radial velocity estimate from 313 to 349.

While a thorough analysis of the integrated properties
(including Lick indices) of the newly confirmed clusters is demanded to Paper
II of this series  (see Sect.~1), a first glance to
their composing stellar populations is provided in Fig.~\ref{jkbk}, where we
compare the new GCs with the already confirmed ones (from G04) and
with the Buzzoni \cite{buz} theoretical models for Simple Stellar Populations (SSPs),
in the reddening-corrected J-K vs. B-K plane (see James et al. \cite{james}
for a thorough discussion of this diagnostic plane). It is clear
that most of the new clusters have the typical integrated colors of classical old
globulars, and they appear to span the whole metallicity range covered
by previously known M~31 GCs.

\begin{figure}
   \centering
   \includegraphics[width=\hsize,clip=]{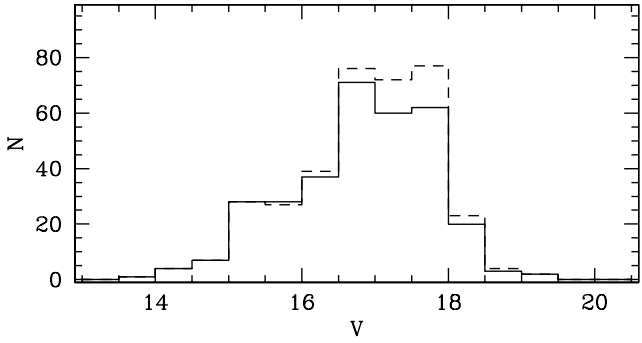}
      \caption{The $V$ luminosity function of confirmed M~31 clusters.
      Continuous and dashed lines show the
      distributions {\em before} and {\em after} the present survey,
      respectively. Most of the new additions (15 out of 42)
      are for class D candidates according to the Battistini et al. classification scheme.
               }
      \label{sur}
   \end{figure}

In terms of GC luminosity function, our survey provides a strong contribution in
the $17.0<V\le 18.0$ range, where we add 30 new clusters to the 120 previously known
objects (see Fig.~\ref{sur}).
The 42 newly confirmed clusters are distributed along all of the Battistini
at al. classes from A to D,\footnote{Specifically, the quality-class distribution results
[A: B: C: D: E]~=~[13: 10: 4: 15: 0].} but we remarkably increased (+75\%) the
surveyed fraction of class D candidates.

 It is important to consider that
only 172 candidates of class D and E have been scrutinized to date.
Since 67 of them turn out to be genuine clusters ($\sim$35\%) and $\sim$500
candidates of these classes still need to be confirmed, one might conclude
that over 100 genuine clusters are still hidden in this  harvest of
``intermediate/low-quality'' targets. Hence, large surveys,
as  the one presented in this paper, are badly needed to eventually reach a truly
complete sample of M~31 clusters.

%

\subsection{A basic application: the M~31 mass estimate}

A preliminary application of the basic kinematical properties of our enlarged
sample provides results in excellent agreement with the more specific analysis
by P02. The mean systemic velocity of the M~31 GC system (after a 2-$\sigma$ clipping
procedure) is $\langle V_{GC}\rangle = -296 \pm 12$~km~s$^{-1}$ , the median is
$-297\pm 14$~km~s$^{-1}$. The overall velocity dispersion is $\sigma=158 \pm 10$~km~s$^{-1}$.

\begin{figure}
   \centering
   \includegraphics[width=\hsize]{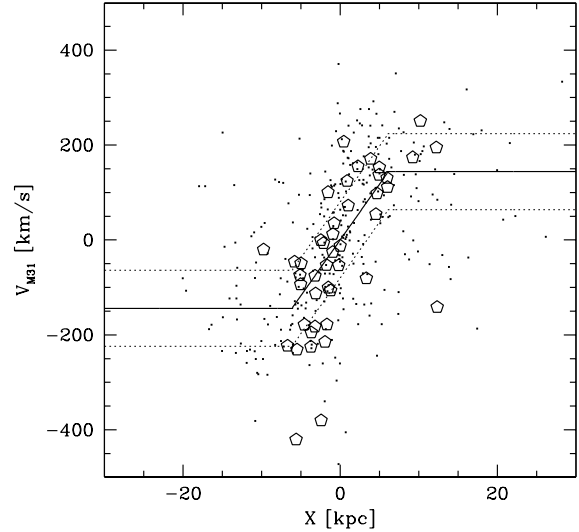}
      \caption{Radial velocity of the M31 globular clusters
      (corrected for the systemic velocity of M31) vs. the projected
      distance along the major axis (X). Pentagon markers represent the newly 
      confirmed clusters studied in the present paper.
      The continuous line is a fit to the 2-$\sigma$ clipped mean velocity as a
      function of major-axis projected distance, computed on 4 kpc wide boxes
      shifted by 1 kpc steps in X.  2-$\sigma$ contours are plotted as dotted
      lines.
           }
      \label{rot}
   \end{figure}

In Fig.~\ref{rot} we show the distribution of the M31-centric velocities of
confirmed GCs as a function of the projected distance from the center along the
major axis of the galaxy (X, see G04, and references therein). To obtain a more
easily readable plot we limited to the range $-30$~kpc $< X < +30$~kpc, while
the outermost cluster in our catalog (B514, see G00) lies at
$X=59.6$~kpc.\footnote{For M~31 we adopt a distance modulus $(m-M)_0=27.47$, from McConnachie et
al. \cite{mcc}, and E(B-V)$=0.11$, as in G04. This corresponds to a distance
$D=783$~kpc.}

The well known rotation pattern of M31 GCs (see van den Bergh \cite{sydbook},
P02, and references therein) is clearly visible in Fig.~\ref{rot},
with a flattening occurring for
$|X|\gtrsim 7$~kpc. Averaging and 2-$\sigma$ clipping velocities as a function of
X in boxes 4 kpc wide, we obtain an amplitude of the overall rotation pattern
of $134 \pm 15$~km~s$^{-1}$, again in very good agreement with P02.

As an example of the possible applications of the newly obtained large and
homogeneous database of radial velocities of M31 GCs, we obtained a simple
estimate of the mass of M31 within $R\simeq 60$ Kpc, using the Projected
Mass Estimator (PME, Bahcall \& Tremaine \cite{bac}.
Adopting the version by Heisler et al. \cite{heis} of the PME
 \begin{equation}\label{eq:massapro}
M_{P}=\frac{C}{\pi GN} \sum_{i}^N V_{i}^2 r_{i}
  \end{equation}
and assuming an isotropic velocity distribution ($C = 32$), P02 obtained for M~31 a
total mass $M_{\rm tot} = 4.1\pm 0.1 \times 10^{11}$~M$_{\odot}$, by using 319
dynamical probes (GCs) out to a radius of $\simeq 27$ kpc from the galaxy
center. Under the same assumptions, from our enlarged sample of 349 GCs out
to $\sim$60~kpc from the galaxy center (projected distance, $R_p$), we obtain
$M_{\rm tot} = 4.4 \pm 0.2 \times 10^{11}$~M$_{\odot}$. It has to be noted that the
uncertainties on the actual isotropy degree of
the underlying velocity distribution contributes an additional factor $\sim$2
to the quoted uncertainties. Note also that the estimate is unchanged if we exclude
from the sample the presumably young clusters identified in Fusi Pecci et al.
\cite{ffp} as belonging to the thin disc of M31. The obtained value is well within the range
spanned by the most recent estimates of the mass of M~31, as listed by Evans \&
Wilkinson \cite{ew00} in their Table~6.
The agreement with previous estimates obtained using GCs as tracers
(Federici et al \cite{luciana}, P02) is also quite good.


%

The Bahcall \& Tremaine \cite{bac} PME  was originally conceived for
test particles to probe a central point-mass gravitational source.
As an illustrative exercise, we can suitably approach this condition
by restraining our analysis to the 14 most distant clusters in our
sample, with  $R_p>20.0$~kpc, and maintain the isotropy hypothesis
(i.e.\ $C = 16$, in eq.~\ref{eq:massapro} above). With these
constraints we obtain $M_{\rm tot} = 4.3-7.0 \times
10^{11}$~M$_{\odot}$, where the reported range has been obtained by
a {\em jackknife} resampling technique (Lupton \cite{lup}). This
estimate is in good agreement with the results by Federici et al.
\cite{luciana} and with the recent independent estimates by Carignan
et al. \cite{cari} and Chapman et al. \cite{chapman}.

The above estimates rely on methods that assume that the adopted tracers 
follow the mass density profile of the probed potential, that is not the case for GCs systems,
in general. To overcome this problem Evans et al. \cite{evans} introduced  a new mass
estimator that doesn't require any coupling between the distribution of the tracers and the
underlying mass distribution. As an example of application, Evans et al. (hereafter E03) 
provide an estimate of the mass of M31 based on GCs, obtaining 
$M=1.2 \times 10^{12}$~M$_{\odot}$. With the same assumptions as Evans et al.
we obtain $M_{rot}=2.9 \times 10^{11}$~M$_{\odot}$ for the
rotational component (to compare with $M_{rot}=3.0 \times 10^{11}$~M$_{\odot}$ found by E03)
and $M_{pres}=2.1 \times 10^{12}$~M$_{\odot}$ for the pressure component
(to compare with $M_{pres}=0.9 \times 10^{12}$~M$_{\odot}$ found by E03). The total mass
is $M=2.4 \times 10^{12}$~M$_{\odot}$, a factor 2 larger than the E03 estimate. Part of this
difference is due to the larger value of the M31 distance adopted here with respect to E03.
Our experiments, however, indicates that the results from this method are quite sensitive to
the the way in which the rotational and pressure components are disentangled: the very simple
rotation pattern adopted here (see Fig.~\ref{rot}) introduces a large uncertainty in our
result. Once the above factors are taken into account the agreement with E03 is satisfying, at
least in this preliminary stage of the analysis.

A detailed analysis of the kinematics of the M~31 GC system and the
galaxy mass profile is deferred to the completion of our
remote-cluster search, currently in progress (see Galleti et al.
\cite{b514}).

\section{Summary and conclusions}

We have presented the first results of a large spectroscopic and imaging survey
of  candidate clusters in M~31. The survey allowed us to confidently classify 76
candidates whose nature was previously unknown: 42 of these resulted new
M~31 GCs, while 12 have been recognized as background galaxies, 2 are M~31 H{\sc ii}
regions, while the remaining 17 objects are foreground stars and 3 unclassified objects
(possibly M~31 clusters or foreground stars). An estimate of the
radial velocity has been obtained for all the 42 newly recognized clusters as
well as for an additional sample of 55 M~31 GCs previously confirmed by other authors
and two controversial objects.
The various set of radial velocities for M~31 GCs available in the literature has
been reported to the same scale, multiple measures have been averaged (see
Sect.~3.2 for details), and a final merged catalog has been produced (see
{\em Tab.~1, Online material}).
The present analysis has increased the sample of confirmed M~31 cluster from
337 to 369 members, and the number of confirmed GCs with a
radial velocity estimate is increased from 313 to 349.

While the main basis for the classification work was provided by radial
velocities, we have also implemented a method to distinguish between
point and extended sources on low-resolution imaging that allows -- in
many cases -- to disentangle genuine M~31 clusters and MW stars when radial
velocity alone leads to controversial conclusions. We also provide a safe
classification for few candidates not included in our survey based on the inspection
of publicly released high spatial resolution images from the HST general archive.

\subsection{The Revised Bologna Catalog V2.0}

All the present observational material has been consistently implemented
to update the RBC, now available on line in its latest V2.0 release.
Future minor updates of the catalog will be described and commented in the RBC web page
({\tt http://www.bo.astro.it/M31/}) where all the catalog database can be
retrieved as ASCII files.

\begin{acknowledgements}

We acknowledge the TNG, WHT and Loiano staff, and especially Roberto Gualandi and
Romano Corradi, for timely and competent assistance during the observing runs.
We also thank Pierre Leisy, for making his software for WYFFOS data reduction
available to us, and Judith Cohen, for her prompt collaboration in checking the
original B216 observations. This project received partial financial support from
the Italian MIUR, under COFIN grant 2002028935-001, and INAF PRIN/05 1.06.08.03.

\end{acknowledgements}

\end{document}